\documentclass[twocolumn]{aastex631}
\usepackage{graphicx}
\usepackage{natbib}

\def\mos{$\mathrm{ms}^{-1}$}

\newcommand{\angstrom}{\mbox{\normalfont\AA}}

\begin{document}

\title{Characterizing and Mitigating Telluric Absorption in Precise Radial Velocities II: A Study of an M2 Type Star}

\author[0000-0001-8079-1882]{Natasha Latouf}
\affiliation{Department of Physics and Astronomy, George Mason University, 4400 University Drive MS 3F3, Fairfax, VA, 22030, USA}

\author[0000-0002-6937-9034]{Sharon Xuesong Wang}
\affiliation{Department of Astronomy, Tsinghua University, 30 Shuangqing Rd, Haidian District, Beijing, China }

\author[0000-0002-2078-6536]{Bryson Cale}
\affiliation{Department of Physics and Astronomy, George Mason University, 4400 University Drive MS 3F3, Fairfax, VA, 22030, USA}

\author[0000-0002-8864-1667]{Peter Plavchan} 
\affiliation{Department of Physics and Astronomy, George Mason University, 4400 University Drive MS 3F3, Fairfax, VA, 22030, USA}

\correspondingauthor{Natasha Latouf, Sharon Xuesong Wang}
\email{nlatouf@gmu.edu, sharonw@tsinghua.edu.cn}

\begin{abstract}
Telluric absorption lines impact measuring precise radial velocities (RVs) from ground-based, high-resolution spectrographs. In this paper, we simulate the dependence of this impact on stellar spectral type and extend the work of the first paper in this series, which studied a G type star, to a synthetic M dwarf star. We quantify the bias in precise RV measurements in the visible and near-infrared (NIR) from the presence of tellurics in a simulated set of observations. We find that M dwarf RVs are more impacted by tellurics compared to G type stars. Specifically, for an M dwarf star, tellurics can induce RV errors of up to 16 cm/s in the red-optical and in excess of 220 cm/s in the NIR. For a G dwarf, comparable RV systematics are 3 cm/s in the red optical and 240 cm/s in the NIR. We attribute this relative increase for M dwarfs stars to the increased concordance in wavelength between telluric lines and stellar Doppler information content. We compare the results of our simulation to data collected on Barnard's star from the iSHELL spectrograph at the NASA Infrared Telescope Facility (IRTF). This study was conducted as a follow-up to the NASA probe mission concept study \textit{EarthFinder}.
\end{abstract}

\section{Introduction}
\label{sec:intro}

Doppler spectroscopy is currently the most prolific method for measuring the masses and characterizing the orbits of exoplanets. Many factors contribute to the measured radial velocity (RV) precision, including limitations from photon noise, instrument precision, telluric absorption, and stellar activity \citep{plavchan2015radial,fischer2016}. New precise radial velocity (PRV) spectrometers such as ESPRESSO, NEID, MAROON-X and EXPRES now have the requisite instrument stability to be sensitive to habitable zone (HZ) Earth-mass planets \citep{expres, neid, espresso,maroonx}. Additionally, these spectrometers have broadened their spectral grasps into the red-optical out to $\sim$900 nm, and several near-infrared (NIR) PRV spectrometers have recently demonstrated $<$3 m/s long-term RV precision at wavelengths out to 2.4 $\mu$m, such as IRD, HPF, CARMENES, SPIRou, and iSHELL \citep{ird, hpf, carmenes, spirou,cale2019}. 

\begin{figure*}[t]
\centering
\includegraphics[width=1.\textwidth]{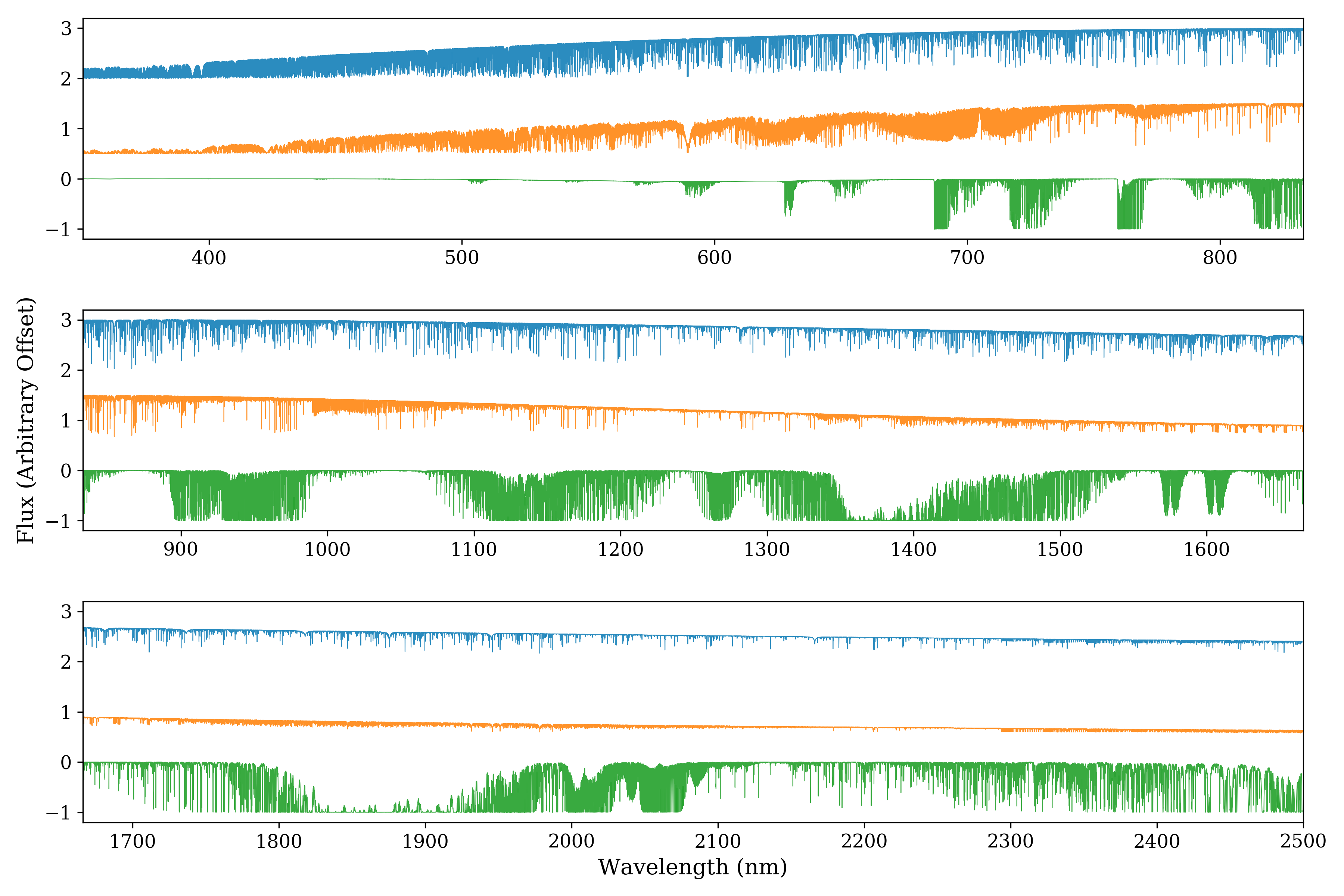}
\caption{Top: Fluxes with arbitrary offsets of the Kurucz solar model, BT-Settl M dwarf model (\citet{allard2012a, allard2012b}), and normalized telluric spectra gathered from TAPAS (\citet{tapas}), shown in blue, orange, and green, respectively, plotted as a function of wavelength in nanometers. This plot ranges from 350 - 850 nm. There is more spectral information content in the M dwarf red-optical spectrum, where there are more telluric lines. By contrast, the solar spectrum has more information content in the optical, with less telluric lines. Middle: The same as the top panel, but from 850 - 1650 nm. Bottom: The same as the other two panels, but from 1650 - 2500 nm.
\label{fig:spectel}}
\end{figure*}

Stellar activity challenges our current sensitivity to HZ Earth-mass planets, the latter of which produce $\sim$8-25 cm/s Doppler reflex semi-amplitudes for G type to M dwarf stars.
By comparison, spots, plages/faculae, the suppression of convective blueshift, and other photospheric effects induce apparent RV changes on the order of a few m/s \citep{dumusque2011, hayword2020, cegla2019stars, dumusque2018stars}. Stellar activity is inherently chromatic, and can be larger in amplitude in the visible wavelengths than in the red-optical and NIR \citep{marchwinski2015, robertson2020, plavchan2020earthfinder, tran2021, cale2021diving}. Investigating the impact of stellar activity on PRVs as a function of wavelength may help to isolate stellar activity from Keplerian (achromatic) signals. Increased spectral grasp of PRV spectrometers have also enabled more RV searches for exoplanets orbiting M dwarf stars, especially the cooler, fainter, late types \citep{morales2019}. Unlike G dwarfs, M dwarfs have the highest RV information content in the red-optical and are brightest in the NIR \citep{reiners2018,reiners2010}. M dwarfs stars are also the most abundant main sequence stars and exoplanet hosts \citep{henry2018, benedict2016, dressing2013,raghaven2010}.

Consequently, we are motivated by the chromaticity of stellar activity and RV information content for G and M dwarfs alike to investigate the challenges to RV precision across a broad spectral range. At almost all wavelengths, tellurics are present in the spectra recorded with ground-based PRV spectrometers. At visible wavelengths, which we define in this work to be 400-700 nm, telluric lines appear to be limited within certain oxygen bands, but in reality, there are copious micro-telluric lines. These shallow lines, typically ignored by telluric masks when extracting radial velocities (e.g., see \citealt{cunha2014}), are dominated by water, a molecule with a strong electric dipole moment, producing absorption features that are small in depth ($\sim$1\%) but highly variable with time and airmass. At red-optical wavelengths, which we define to be $\sim$700-1000 nm, both micro-telluric and deeper macro-telluric lines are present from water and molecular oxygen (O$_2$). At near-infrared wavelengths, which we define to be $\sim$1-2.5 $\mu$m, more prevalent and deeper telluric absorption lines are formed from water in addition to methane, carbon dioxide, nitrous dioxide, and oxygen, producing lines that can be saturated and remain variable with time and airmass. These lines are challenging to model to an accuracy of  1-2\% in flux, \citep[e.g.,][]{smette2015, gullikson2014, seifahrt2010, ulmer-moll2019}. However, when analyzing real RV observations, micro-tellurics are typically comparable to or below the noise level and are either masked or ignored altogether.

In this work we specifically consider the impact tellurics have on PRV measurements of M dwarfs. This project is a direct continuation from \citet[][hereafter W22]{Wang2022}. In W22, we quantified the error due to tellurics for a solar twin star (G2V), using the Kurucz solar model \citep{kurucz2005}. Using simulated spectra, we extracted the RVs and assessed the RV precision as set by telluric contamination alone. By generating noiseless simulated spectra, we isolated the telluric impact on PRV measurements from other sources such as photon noise and stellar jitter. We extracted the RVs with the cross-correlation function (CCF) and the Forward Modeling method in a broad wavelength range (350 - 2500 nm). 

In this project, we follow the same methodology as in W22, with the substitution of a synthetic M dwarf model instead of the solar model. There is significant wavelength overlap between the stellar RV information content and the telluric absorption lines for M dwarfs, and thus we are motivated to quantify the RV error budget term due to the telluric absorption. In $[\S]$ \ref{sec:method} we present the methodology of our experiment, also providing a brief summary of W22. In $[\S]$ \ref{sec:results}, we present the results of our experiment with the M dwarf stellar model. In ${\S}$ \ref{sec:disc}, we analyze the differences between the G type star results presented in W22 with the results of the M dwarf contained herein. We present a comparison to our simulation of an M dwarf with an analysis of iSHELL data of Barnard's star in subsection \ref{sec:ishell}. We present our conclusions and future work in $[\S]$ \ref{sec:conc&fw}.

\section{Methodology}
\label{sec:method}

In this section, we describe the simulations used in our analyses. We follow a similar approach to W22. Herein we present a summary of the main steps in our analysis. Our simulation's code is summarized in a flow chart in W22. The code uses several inputs, including:
\begin{itemize}
\item{A model for the stellar M dwarf spectrum.  We use a BT-Settl model, rotationally broadened to a $v$sin$i=2$ km/s, which is representative of a relatively quiescent M dwarf \citep{allard2012a, allard2012b}}.
\begin{itemize}
    \item{We used the BT-Settl atmospheric models from \citet{allard2012a,allard2012b, baraffe2015} to generate the synthetic spectrum, with version 15.5 of the PHOENIX stellar atmosphere modeling code. The spectrum has been continuum normalized by fitting a blackbody curve initially and then refining it by using a power-law to continuum normalize certain sections that were not well normalized in the first pass.}
    \item{The stellar parameters used to generate the synthetic spectrum are as follows: Effective temperature of 3563K, Metallicity of $[M/H]=-0.38$, $\log{g}=4.8$ to best match the stellar spectrum of the RV standard star HD 95735 as used in \citet{wang2019} (see section 2.1 therein for more detail). This will enable us to compare results in the V band in the optical directly with \citet{wang2019}, which studied the effects of telluric lines in the iodine-calibrated PRVs in the V band, at roughly 500-600 nm.} 
\end{itemize}
\item{A suite of models for the Earth's atmospheric absorption provided by TAPAS, containing data that were computed based on the atmospheric conditions of a random night per month over a year from a location on Kitt Peak, i.e. we generated 12 spectra suites, generally from the 15th day of each month for consistency, unless data was unavailable \citep{tapas}. The TAPAS suite models used for the purposes of this project was identical to the suites used in W22.}
\item{A single Gaussian model for the line spread function (LSF) equivalent to a spectral resolving power of $R=120,000$, which is the approximate spectral resolution commonly adopted by the next generation, ground-based RV instruments, e.g., NEID, EXPRES, ESPRESSO. This is kept unchanged throughout our simulations.}
\item{The airmass and optical depth ($\tau$) for 6 molecular species (water, methane, carbon dioxide, nitrous oxide, oxygen, and ozone molecules), as well as a random value drawn from a distribution of real site values for the amount of precipitable water vapor (PWV). The barycentric velocity correction (BC) was calculated as in W22: 
\begin{equation}
   BC = sin(date/(2\pi \cdot 365.15) - \pi/2) \cdot 30 km/s
\end{equation}
The telluric spectrum was shifted according to the calculated BC value to mimic Earth's motion. We shifted the telluric spectrum rather than the stellar spectrum due to the higher fidelity of TAPAS relative to the fidelity of the stellar spectrum. The barycentric velocity, airmass, and PWV distributions are the same between W22 and this experiment to allow for a direct comparison.}
\end{itemize}

A direct comparison of the Kurucz solar model, the BT-Settl M dwarf model, and the telluric model from TAPAS is shown in Figure~\ref{fig:spectel}. As seen and described further, the solar model has more RV information content in the optical wavelengths, less plagued by tellurics. For the M dwarf, however, the peak RV information content occurs in the red-optical, where there are deeper telluric absorption lines.

We make several assumptions in the setup of our analysis. In particular, we assume we are observing from a telescope on Kitt Peak. We assume that we observe 365 nights per year, once for each night of the year, e.g. a circumpolar best case scenario. As in W22, we compute our observed spectra in the rest frame of the star.  Thus we shift our model telluric spectrum by the negative of the barycenter velocity. We approximate the barycentric velocity correction as a sine wave with a one year period, phase, and semi-amplitude of 10.7 km/s that approximately matches the archival observations of Tau Ceti with Keck/HIRES \citep{howard2016, tuomi2013}, to match with the assumption in W22 since, as can be seen later, the individual RV measurement errors correlate strongly with the barycentric RVs.

For these simulations, we used the BT-Settl M dwarf spectral model. We assume no stellar activity modifies the stellar spectra and line shapes. We then multiplied the M dwarf spectra with the TAPAS atmosphere, which varies by night for each of the 365 nights in our simulations according to the airmass, the PWV, and the barycentric correction (BC) of each night. Lastly, we convolved the spectra at a resolution R = 120,000, as expressed above. We did not add photon noise or order blaze functions, in order to further isolate the errors due to tellurics.\footnote{For more information, please see the discussion on caveats in W22.} Then, we extracted the RVs using two methods: CCF and Forward Modeling. Each of these two methods was run with two sets of assumptions: with perfectly known telluric line profiles and imperfectly known telluric line profiles.

The first method used for extracting RVs was the CCF \citep{baranne1996}:
\begin{enumerate}
    \item{No Correction: In this method, RVs were extracted using the method similar to \cite{zeichmeister2018} using a stellar template instead of a line mask. We also assumed, unrealistically, that we have a perfectly known wavelength solution, LSF, and stellar template. We do not apply any telluric line-masking before extracting RVs, as is done with {\tt{SERVAL}} and other RV analyses with the CCF method at visible wavelengths.}
    \item Correction by Division: In this method, we assume a perfectly known telluric line that is the same as the telluric absorption that we put in for each night. We correct for tellurics by dividing the telluric lines (broadened to a resolution of 120,000) before extracting the RVs. This is the most commonly used method of correction for instruments such as HARPS or CARMENES, which are stabilized spectrographs (e.g., \citealt{cunha2014,artigau2014}). While this method works fairly well, it is also technically mathematically incorrect, involving division after, rather than before, convolution by the LSF. For more information on the incorrect mathematical treatment of dividing an observed spectrum by a telluric model, consult W22.
    \item Correction by Division with Wrong Telluric Line Profile: In this method, we divide out a set of telluric lines generated with a line profile that is mismatched to the telluric line profile we used when multiplying to form our total simulated spectra. Here, the wrong telluric line profile refers to the intrinsic line profiles of the telluric lines and not the spectrograph line spread function. The mismatched telluric profile is based on the atmospheric model of a different observing site than our assumed site of Kitt Peak. Same as in W22, we used Maunakea as our mismatched profile location, hereafter known as the K-profile. We selected Maunakea due to this difference in elevation from our assumed observation site of Kitt Peak, which would mimic the lack of precise knowledge of the telluric line profile and molecular abundances in ground-based observations.
\end{enumerate}
The second method for extracting RVs we used was Forward Modeling:
\begin{enumerate}
    \item Perfectly Known Line Profile: In this method, we expect to get a result of 0 m/s error with our RV extraction. This method is used as a control to reaffirm our analysis yields the injected RVs. These results are as expected and not presented herein.
    \item Imperfectly Known Line Profile: In this method, the Forward Modeling algorithm is performed with imperfect knowledge of the existing telluric line profile, the same as in the method of using CCF with correction by division with K-profile as described above. 
\end{enumerate}
For the RVs of each order, we calculate the overall scatter over time (RMS), which is used as the quantification of the errors induced by the tellurics from this order. The spectrum is divided into 230 orders and this is done for all orders. The spectrum division is conducted to mimic real observations, where the RVs tend to be individually computed for each order independently and combined afterwards. The orders are weighted to account for the Doppler content, and the order sizes are set to linearly increase with increasing wavelength to remain constant in $\delta\lambda/\lambda$. In addition, we take a slice through time, getting RVs from all orders for any given night, and statistically combine them according to the stellar Doppler content and its overall scatter due to tellurics. In this way, the final nightly multi-order RV is calculated. 

\begin{deluxetable*}{cccccc}[!t]
    \tablehead{
    \colhead{Ground-based} & \colhead{No Correction} & \colhead{Division} & \colhead{Division} & \colhead{Modeling} & \colhead{Photon-limited} \\
    \colhead{Instrument} & \colhead{(m/s)} & \colhead{(m/s)} & \colhead{K-Profile (m/s)} & \colhead{K-Profile (m/s)} & \colhead{Precision (m/s)}
    }
    \startdata
      EFE Visible Arm & & & \\ (380-900 nm) & 0.167 &   0.125 &   0.120 &   0.140 &   0.102 \\ \hline
      EFE NIR Arm &&&&\\
      (900-2500 nm) & 2.158 &   0.636 &   1.622 &   1.636 &   0.160\\ \hline
      ESPRESSO &&&&\\(380-788 nm) & 0.166 &   0.137 &   0.119 &   0.137 &   0.109 \\ \hline
      EXPRES &&&&\\(380-680 nm) & 0.149 &   0.042 &   0.119 &   0.104 &   0.138 \\ \hline
      NEID &&&& \\(380-930 nm) & 0.167 &   0.124 &   0.112 &   0.127 &   0.104 \\ \hline
      CARMENES &&&& \\Visible Arm &&&& \\(520-960 nm) & 0.302 &   0.170 &   0.171 &   0.198 &   0.159 \\ \hline
      CARMENES &&&& \\NIR Arm &&&& \\(960-1710 nm) & 2.122 &   0.685 &   1.149 &   1.156 &   0.293 \\ \hline
     \enddata
     \caption{Summary of RV RMS of an M dwarf for different instruments including an EarthFinder ground equivalent instrument using the different RV extraction methods previously discussed in Section \ref{sec:method}. Columns 2-5 remain the same as Columns 3-6 in Table~\ref{tab:resulttabboth}. Column 6: theoretically computed SNR = 100 photon-limited precision (m/s) for our chosen M dwarf in each band.}
     \label{tab:rvinst}
\end{deluxetable*}

\begin{figure*}[t!]
\centering
\includegraphics[scale=0.43]{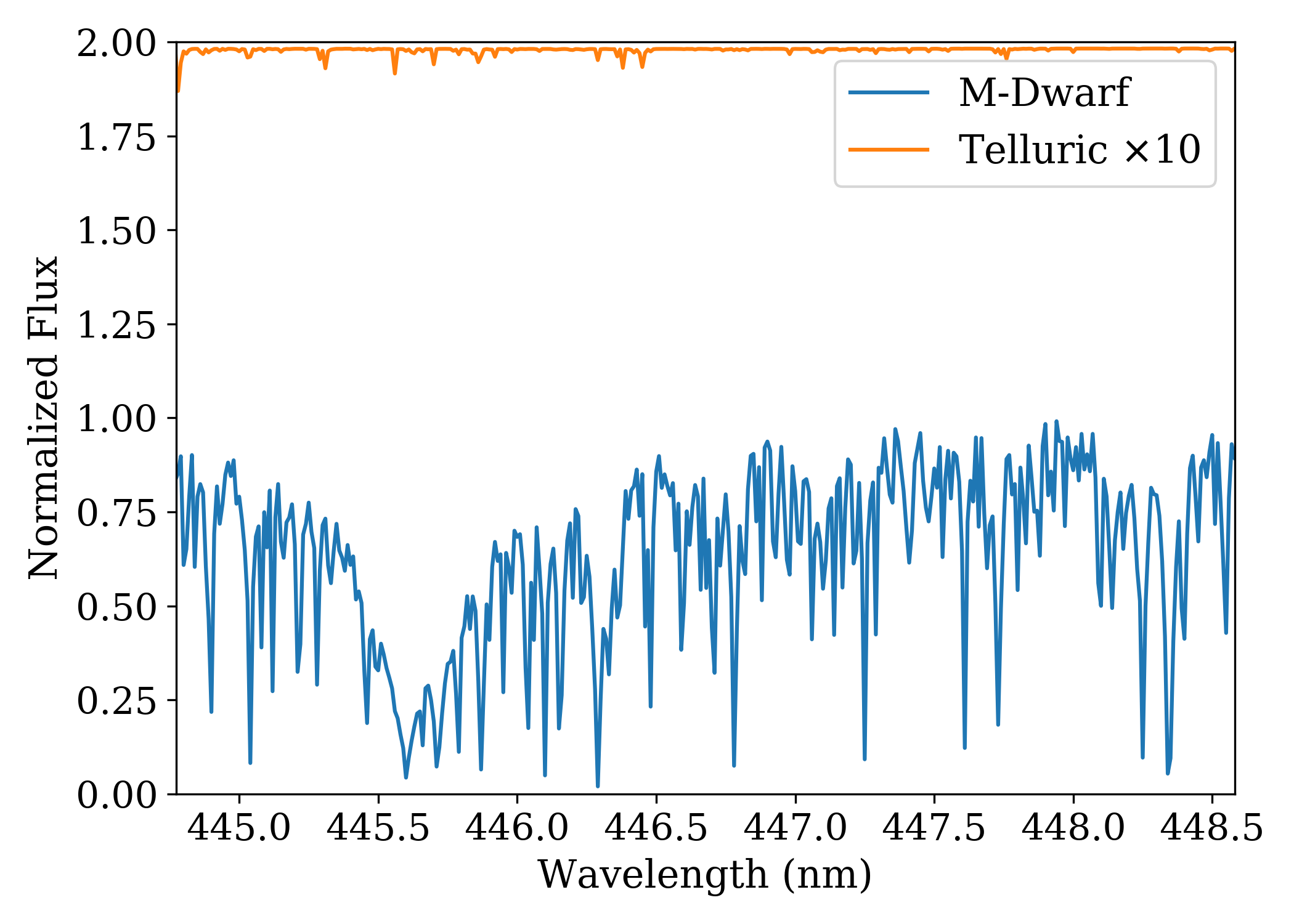}
\includegraphics[scale=0.43]{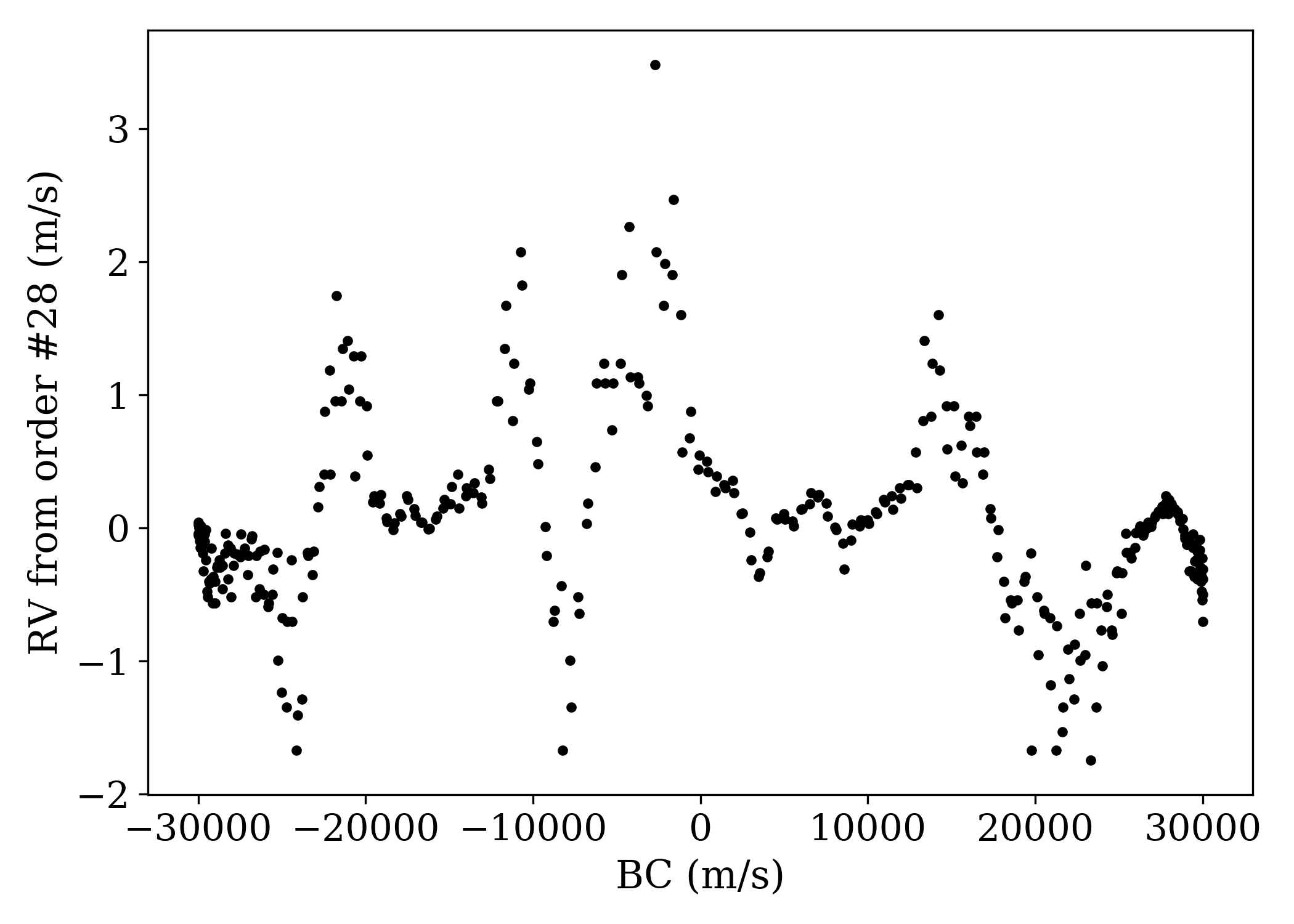} \\
\includegraphics[scale=0.43]{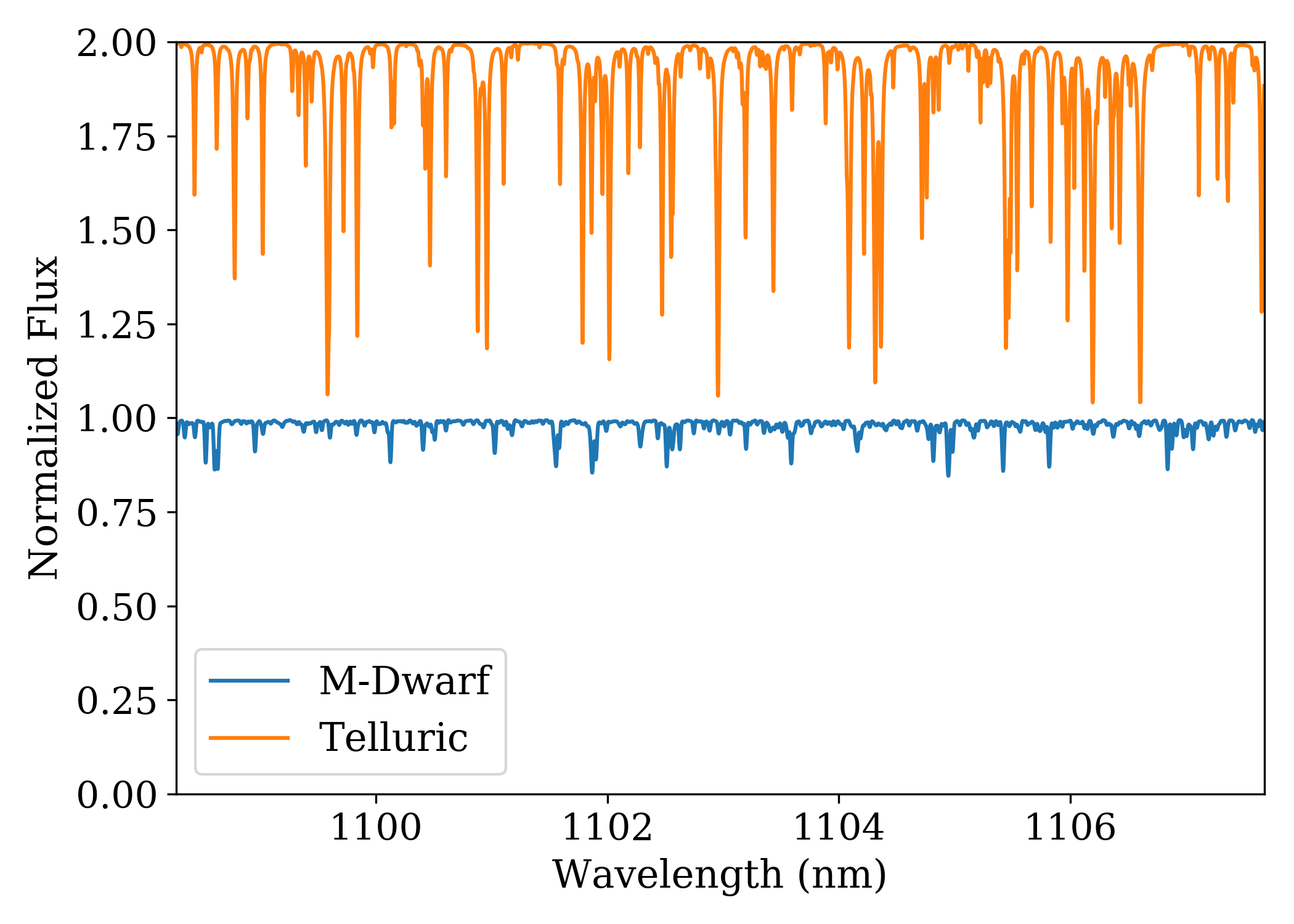}
\includegraphics[scale=0.43]{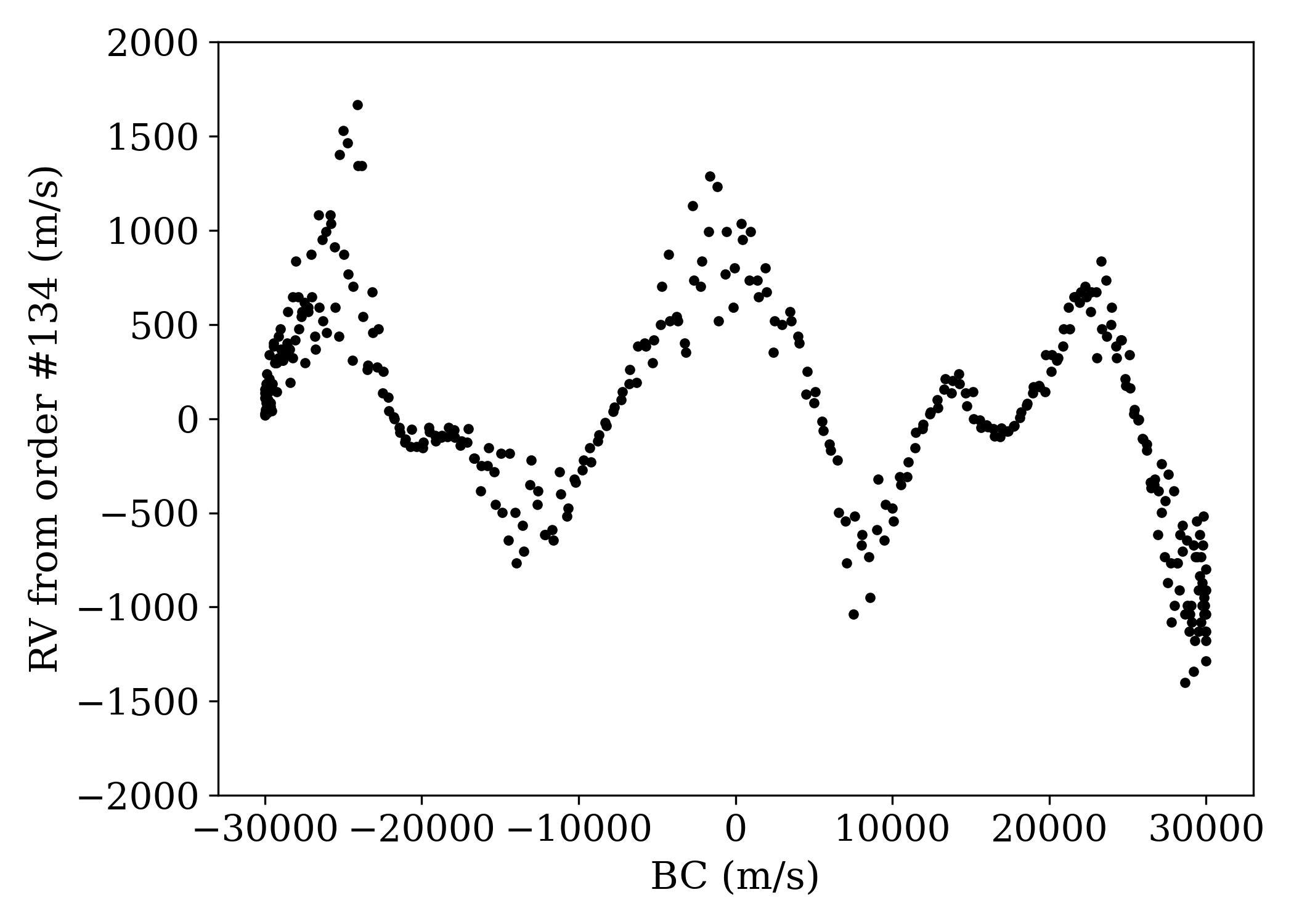}
\caption{Example spectral orders shown on the left with their corresponding RVs vs.\ the barycentric correction on the right when specifically ignoring tellurics during RV extraction, i.e. CCF + no correction. The left plots show the M dwarf spectrum in blue and the telluric spectrum in orange for each order, with the telluric absorption on order \#28, shown in the top left panel amplified by 10 in flux for better visibility. 
}
\label{fig:chunks}
\end{figure*}

\section{Results}
\label{sec:results}

Herein we present our results for the simulation of the BT-Settl M dwarf model, as compared to the prior results of the G star simulation. All plots are replicated in similar forms to W22 for consistency, and comparative material has been added where necessary.

\label{sec:rvcorr}

First, we present example spectral orders with corresponding RV vs barycentric correction in Figure~\ref{fig:chunks}. Looking at the two left panels (top and bottom), we can see the telluric spectrum and M dwarf spectrum between 445 - 448.5 nm and 1099 - 1107 nm, respectively. The corresponding right plots show how varying depths of telluric lines affect RVs to different degrees and that the RV biases caused by tellurics in the temporal space are highly structured due to Earth's barycentric motion. This is caused by the changes in wavelength separation
between the telluric lines and the stellar lines due to the
Earth's barycentric motion.  This rastering is also discussed in W22 and other prior work, e.g. Figure 3 in \citet{cunha2014} and references therein. This shows how tellurics add to the RV error budget, by adding scatter to the measured RVs and adding coherent systematic signal. This is described further in W22, presenting the peak pulling effect and the dependence of amplitude on the relative Doppler content of the stellar and telluric lines.

\begin{figure*}[t]
\centering
\includegraphics[scale=0.38]{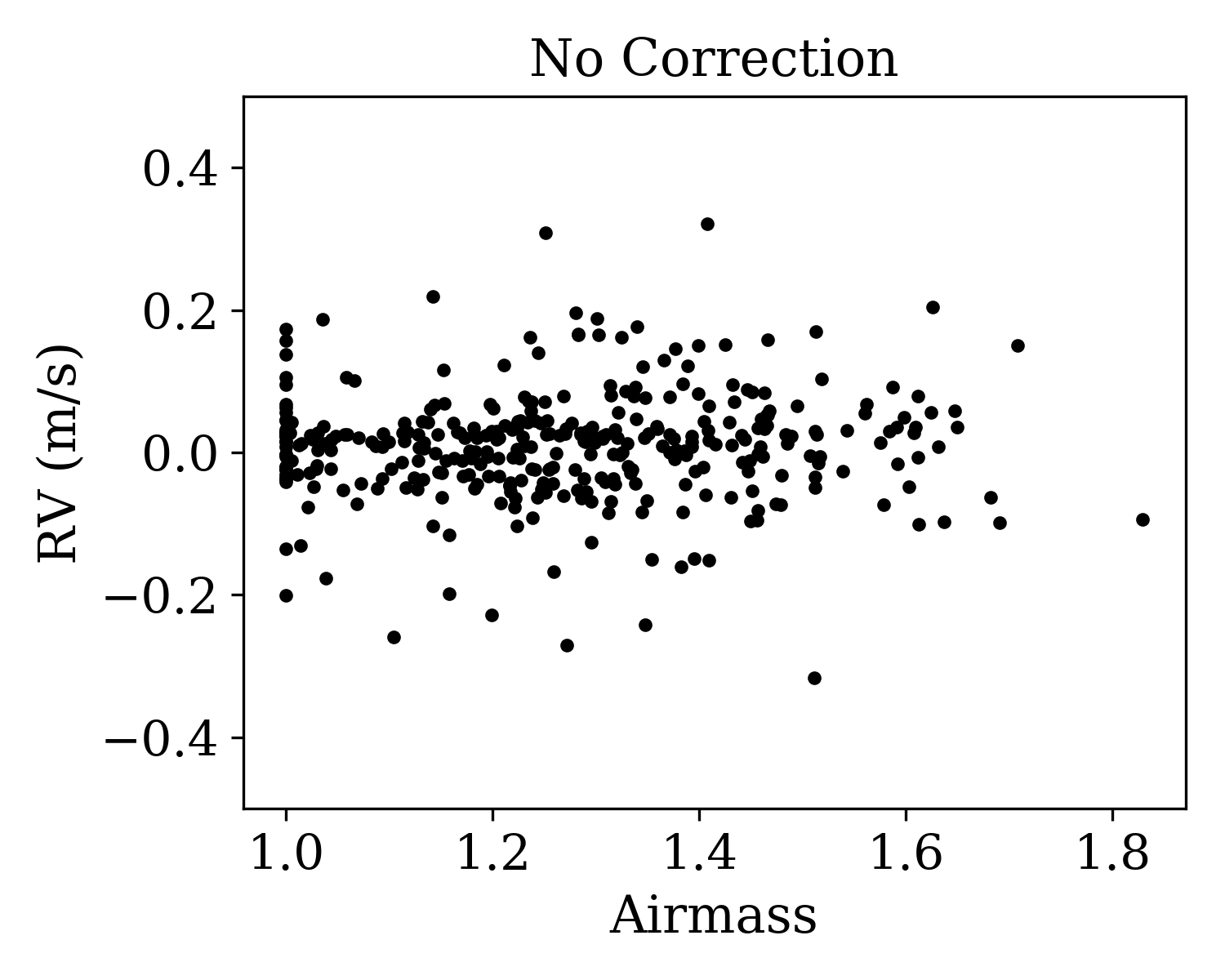}
\includegraphics[scale=0.38]{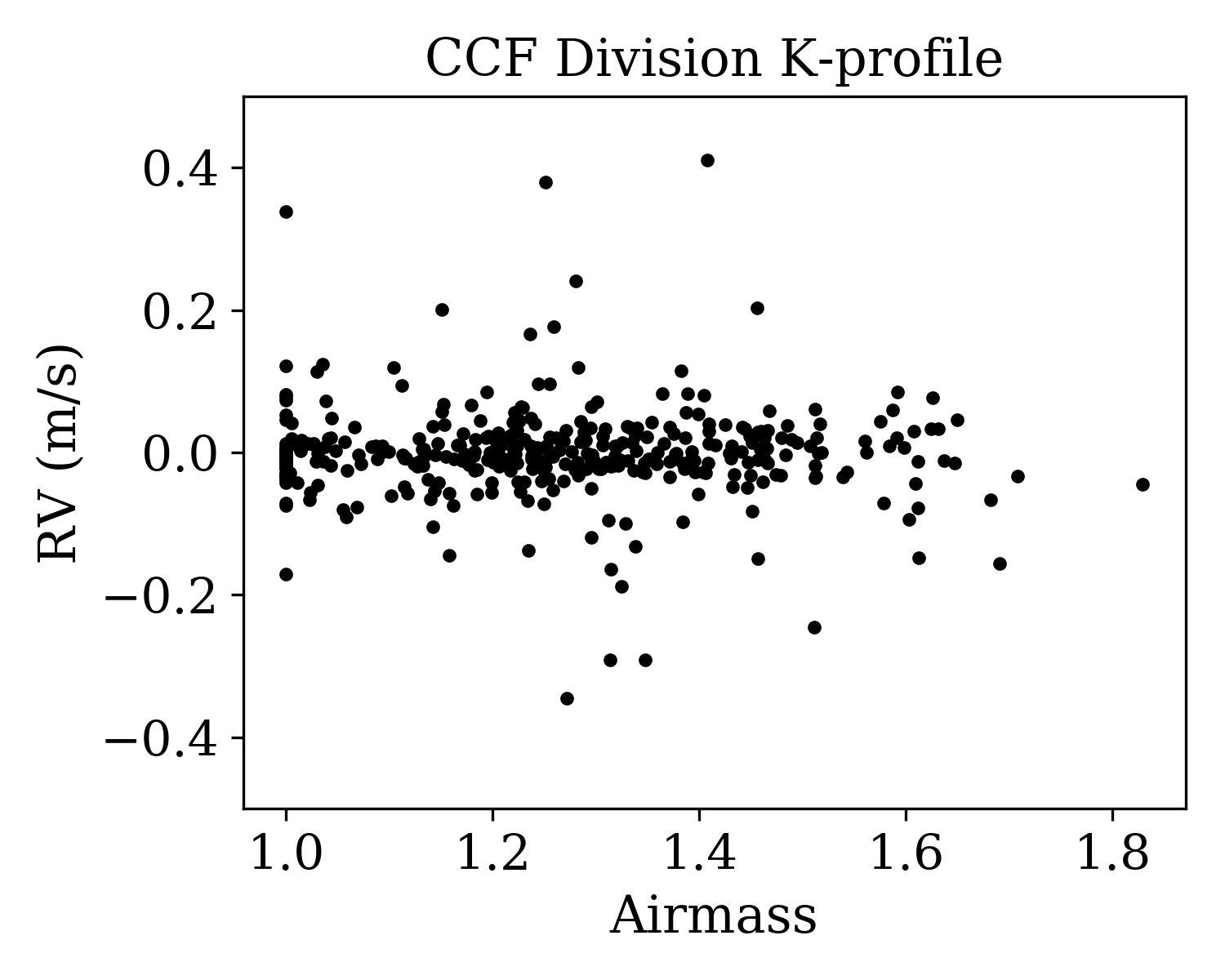}
\includegraphics[scale=0.38]{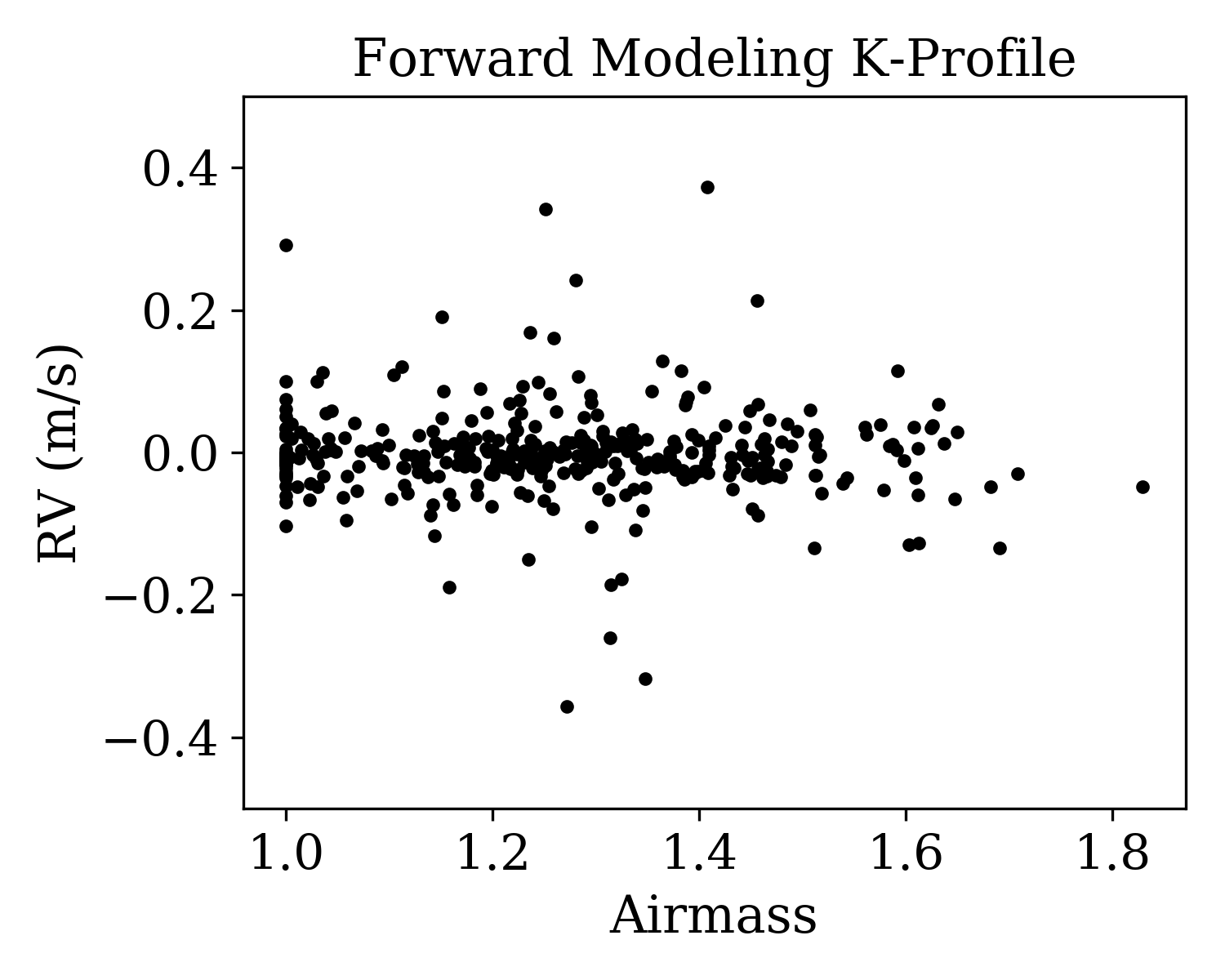} \\
\includegraphics[scale=0.38]{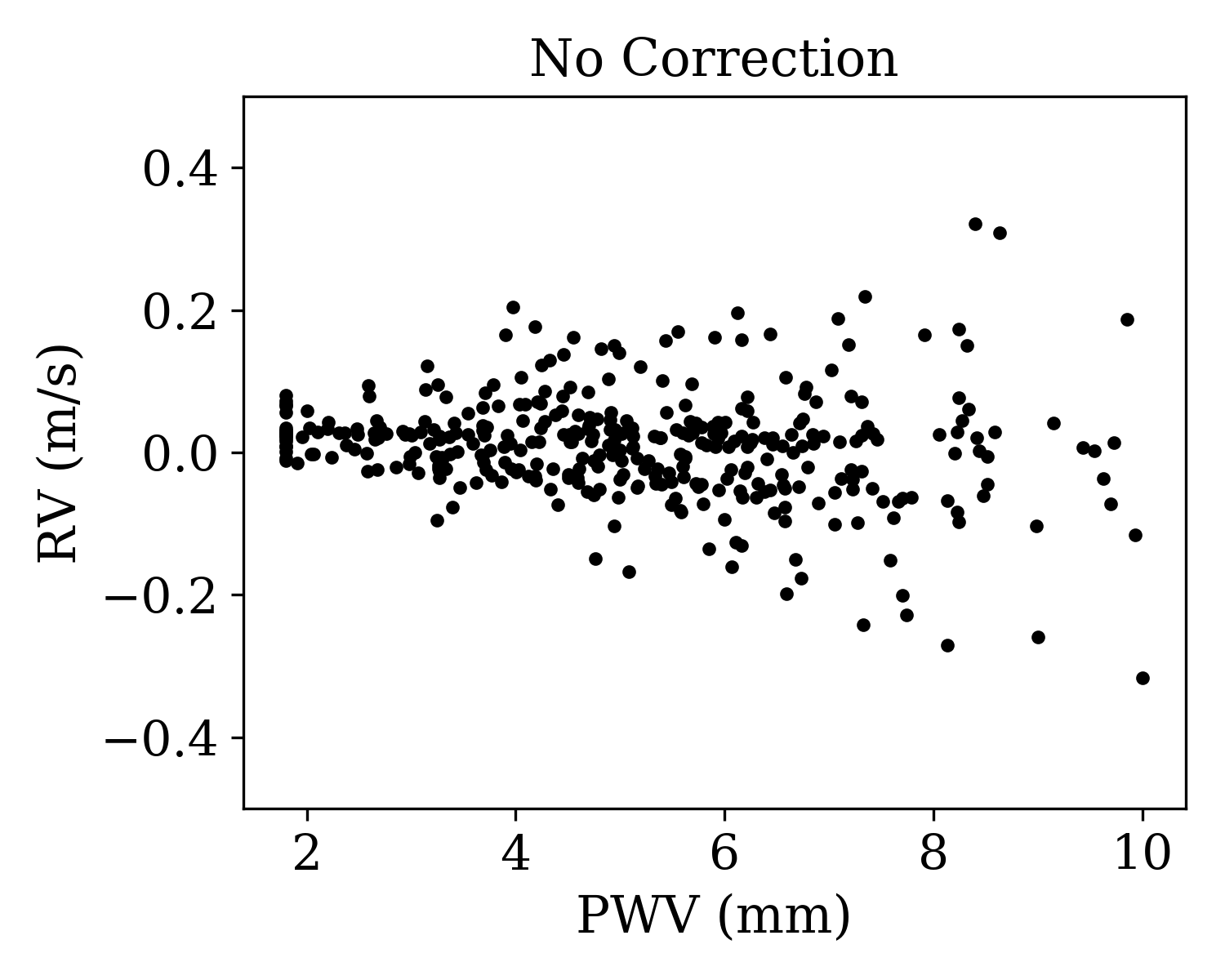} 
\includegraphics[scale=0.38]{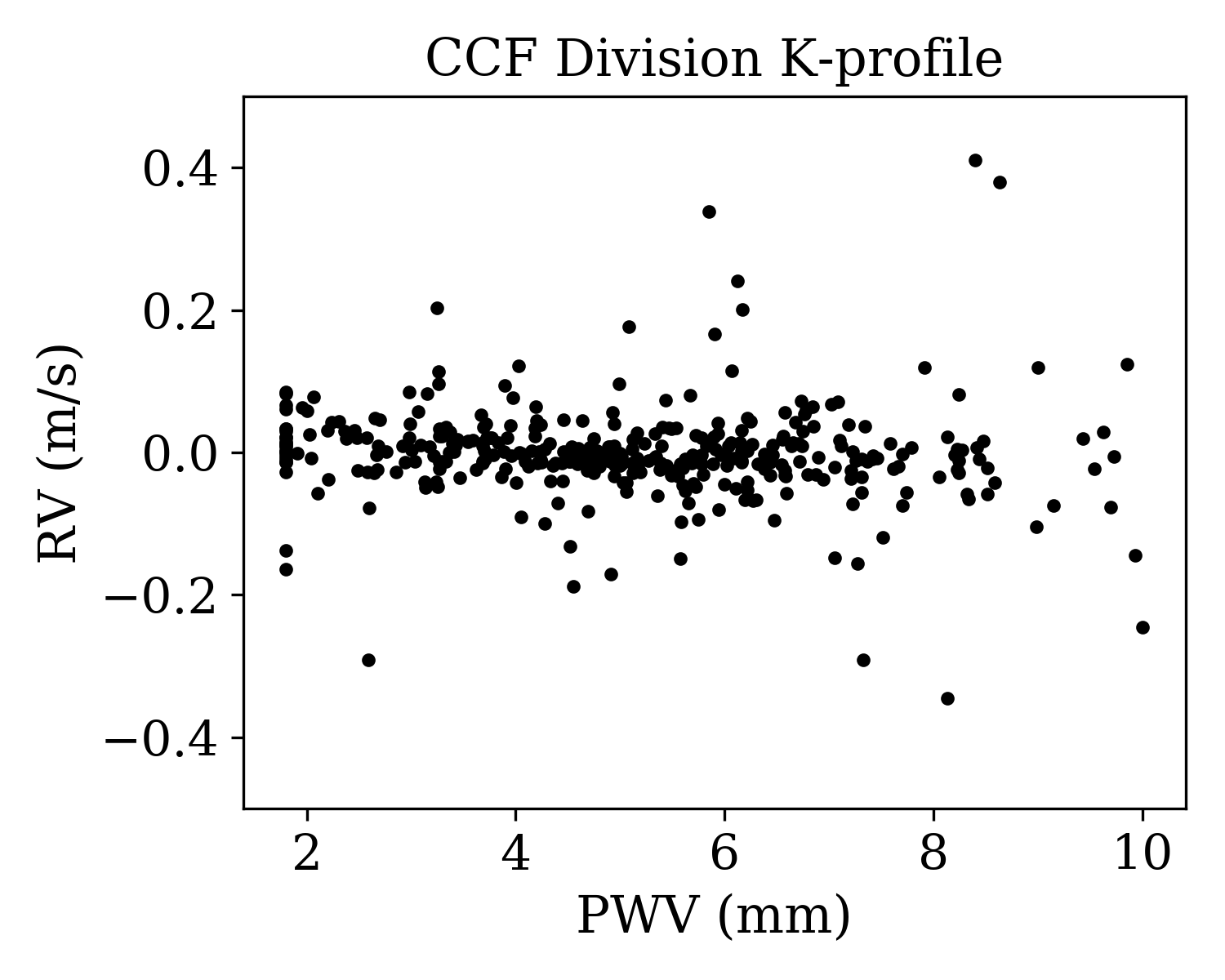} 
\includegraphics[scale=0.38]{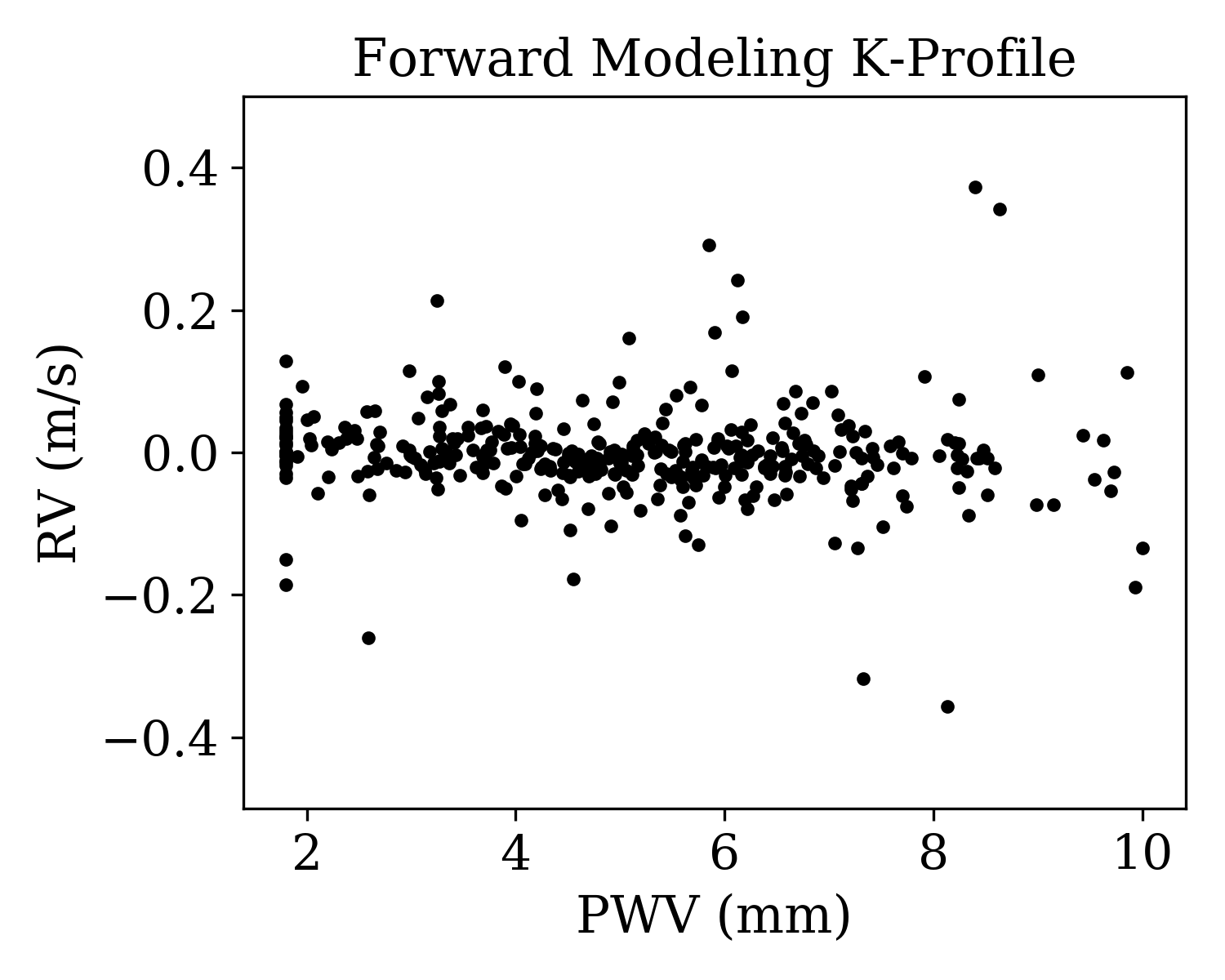} \\
\includegraphics[scale=0.38]{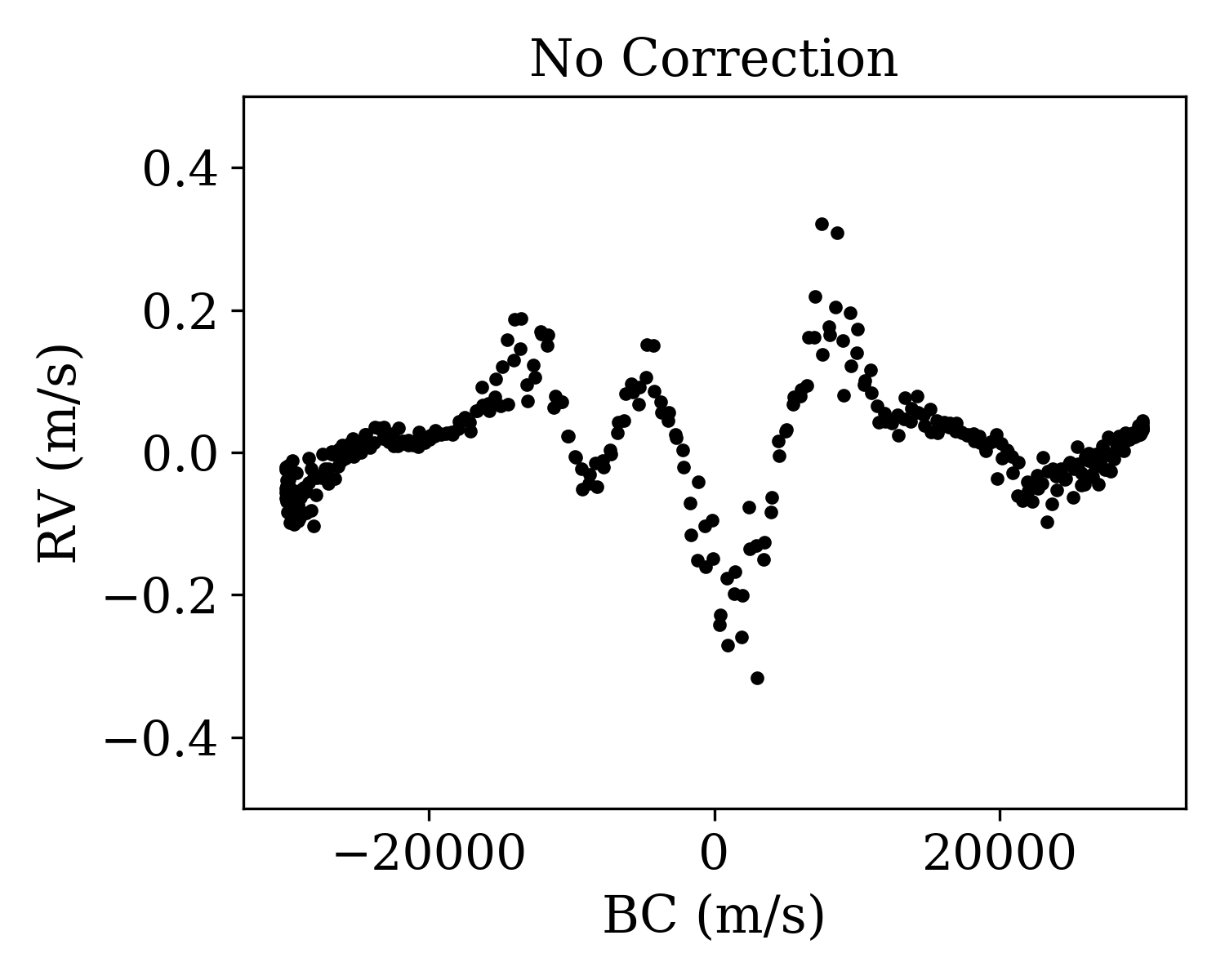} 
\includegraphics[scale=0.38]{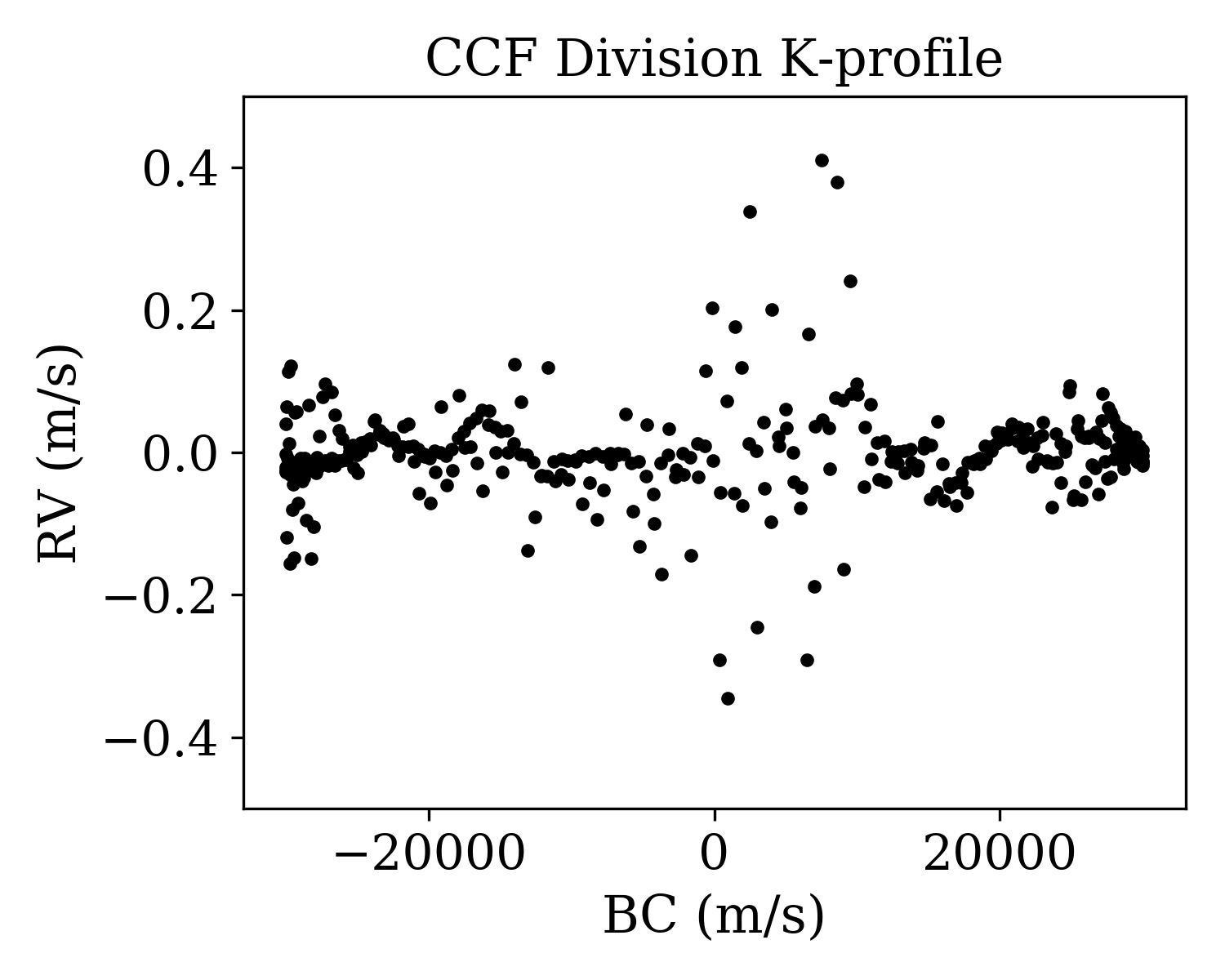} 
\includegraphics[scale=0.38]{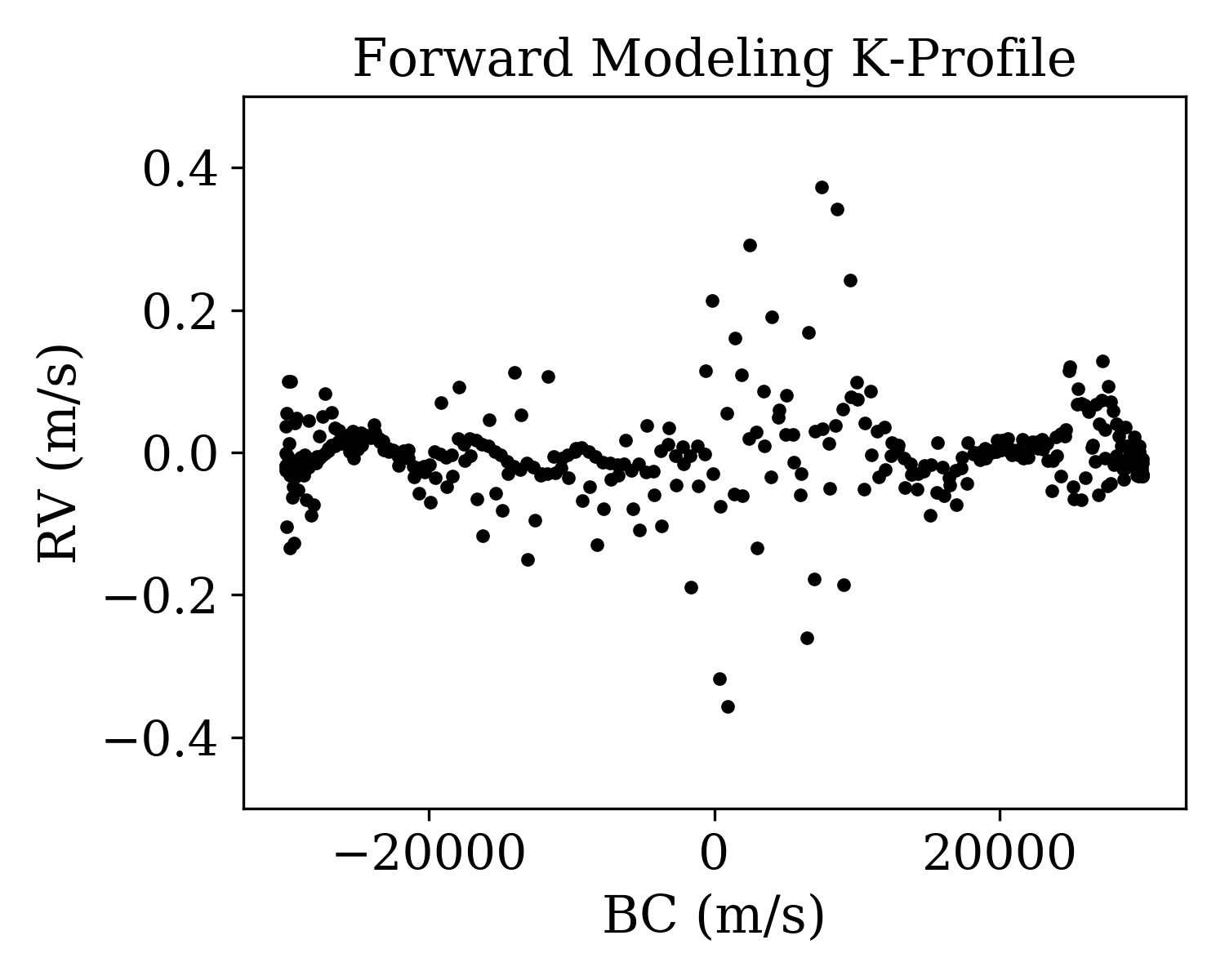} 
\caption{Extracted RVs from 365 nights as functions of airmass, PWV, and BC for three methods: No correction (left), CCF$+$division with the K-profile for tellurics (middle), and Forward Modeling with the K-profile for tellurics (right). Each data point is a weighted average of reported RVs from all orders from 350 - 2500~nm, the entire wavelength range in our simulations. The weights assume SNR = 400 per pixel and combine photon noise with telluric-induced RV RMS, as described in W22, Section~\ref{sec:method}, and as used in the presented data in Table~\ref{tab:rvinst}.
\label{fig:pabcalc}}
\end{figure*}

The full suite of extracted RVs as a function of airmass, PWV, and BC is presented in Figure~\ref{fig:pabcalc}. Each data point presented is a weighted average of reported RVs from 350 - 2500 nm, with assumed SNR of 400 per pixel and combined photon noise with telluric-induced RV RMS. The lack of correlation between RVs and PWVs is expected, as the input PWV value would not match the K-profile PWV value and the water lines are largely mitigated, although not perfectly. This also shows that the scatter in RVs is weakly related to the airmass, due to the final RVs being dominated by RVs in the optical region. However, we can see that the BC is the most strongly correlated to the RV RMS, due to the highly structured systematics.

\begin{deluxetable*}{cccccc}
    \tablehead{
    \colhead{Photometric} & \colhead{No Correction} & \colhead{Division} & \colhead{Division} & \colhead{Modeling} & \colhead{Photon-limited} \\
    \colhead{Band} & \colhead{(m/s)} & \colhead{(m/s)} & \colhead{K-Profile (m/s)} & \colhead{K-Profile (m/s)} & \colhead{Precision (m/s)}
    }
    \startdata
      B (398--492 nm) & 0.066 &   0.000 &   0.050 &   0.058 &   0.451\\ \hline
      V (507--595 nm) & 0.233 &   0.021 &   0.154 &   0.169 &   0.339\\ \hline
      R (589--727 nm) & 0.556 &   0.259 &   0.523 &   0.428 &   0.255\\ \hline
      I (732--881 nm) & 1.523 &   0.525 &   1.513 &   1.551 &   0.632\\ \hline
      Z (0.8--1.0 $\mu$m) & 2.624 &   0.886 &   2.223 &   2.306 &   0.625\\ \hline
      Y (0.96--1.1 $\mu$m) & 2.131 &   0.419 &   2.032 &   1.910 &   0.706\\ \hline
      J (1.1-1.3 $\mu$m) & 17.769 &   2.859 &   11.275 &   9.872 &   1.341\\ \hline
      H (1.5--1.8 $\mu$m) & 6.176 &   1.383 &   3.901 &   3.355 &   0.769\\ \hline
      K (2.0--2.4 $\mu$m) & 33.987 &   1.757 &   22.424 &   6.941 &   1.056\\ \hline
     \enddata
     \caption{Summary of RV RMS of an M dwarf for several photometric bands. Columns 2-6 remain the same as columns 2-6 in Table~\ref{tab:rvinst}. }
     \label{tab:rvband}
\end{deluxetable*}
\begin{figure}[t]
\centering
\includegraphics[width=.9\linewidth]{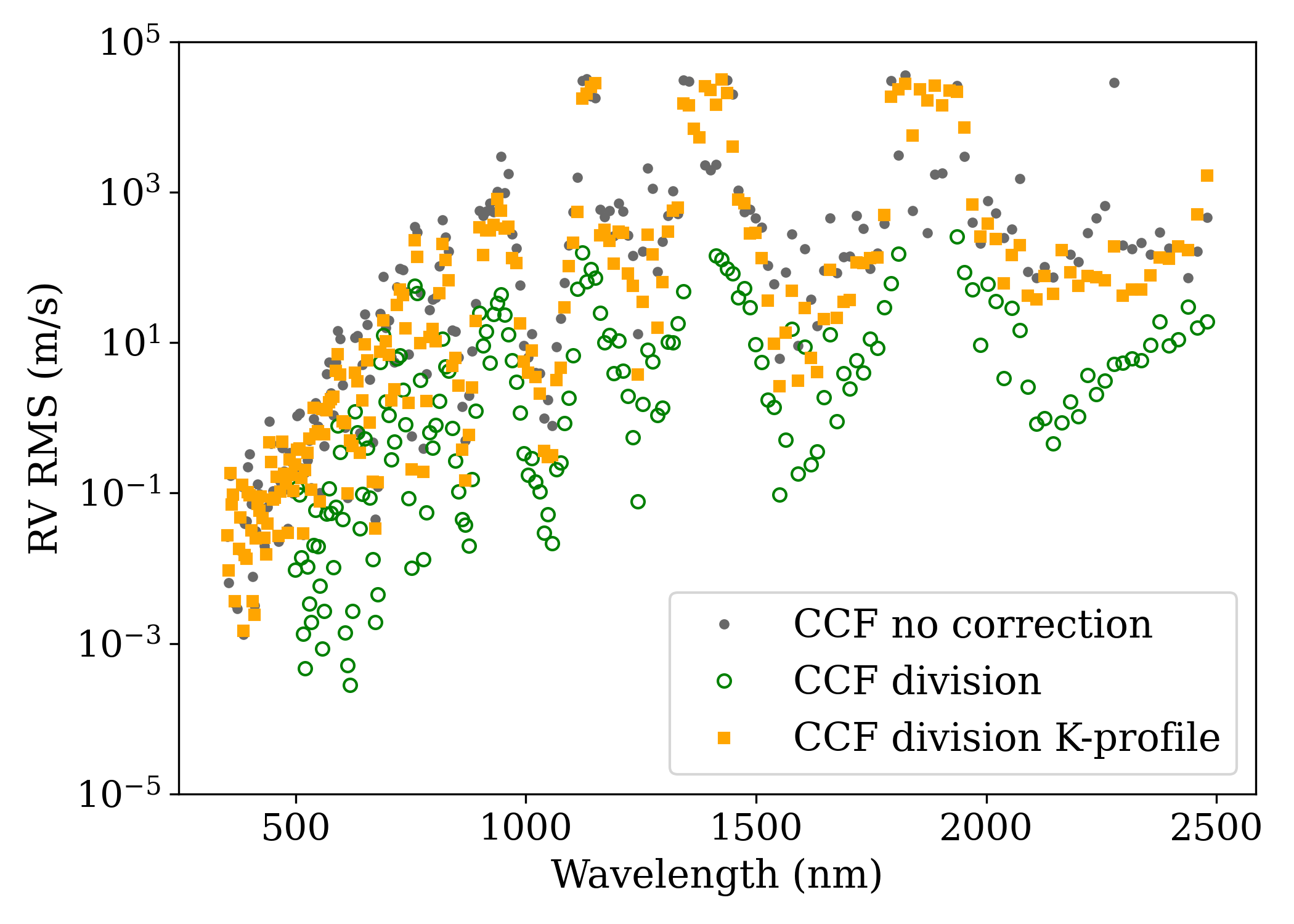}\\
\includegraphics[width=.9\linewidth]{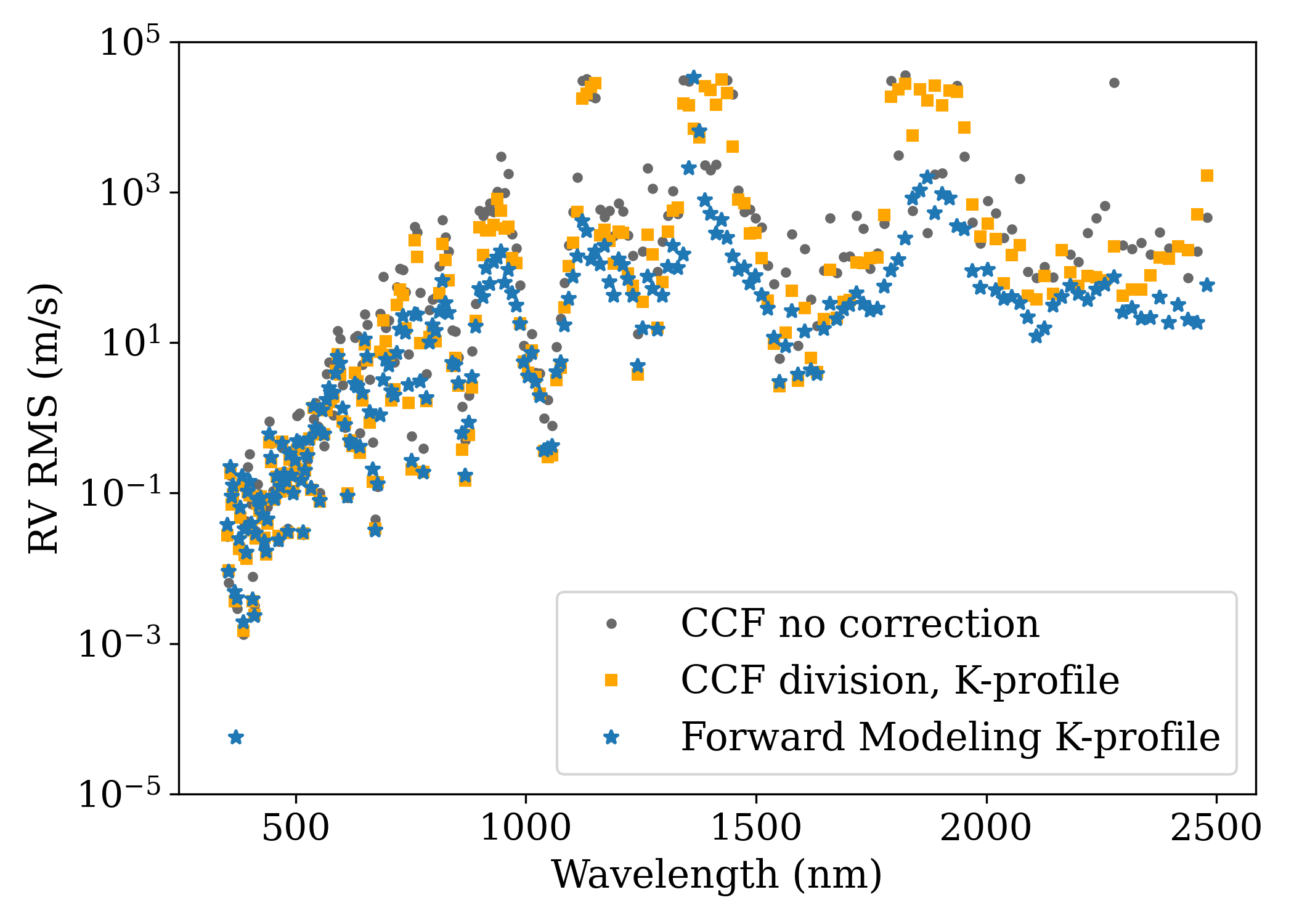}\\ 
\includegraphics[width=.9\linewidth]{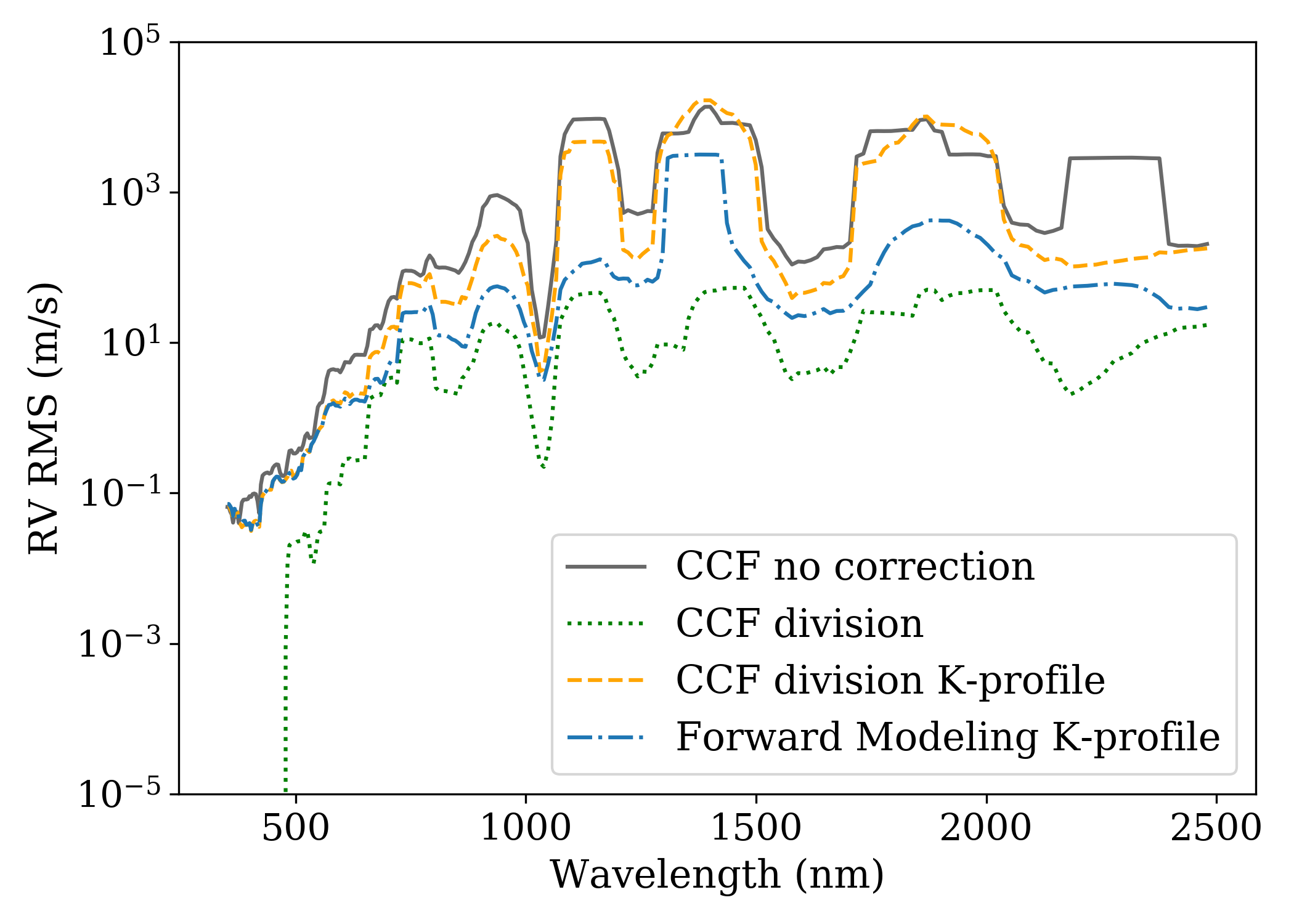}\\
\caption{Telluric line induced RV errors as a function of wavelength for each order in our simulation for the RV extraction methods listed in $[\S]$ \ref{sec:method}. In the top two plots, each point shows the RMS value of RVs extracted from one spectral order each across one year (365 nights). The top left plot shows how using an incorrect line profile introduces more error to the analysis. The top right plot portrays how Forward Modeling generally has a lower RMS than CCF$+$division, especially in the NIR. The bottom plot is the smoothed version of the top plots for more clarity, with a window size of 11 orders and a moving average.
\label{fig:RMSwave}}
\end{figure}

\begin{figure}[t]
\centering
\includegraphics[width=\linewidth]{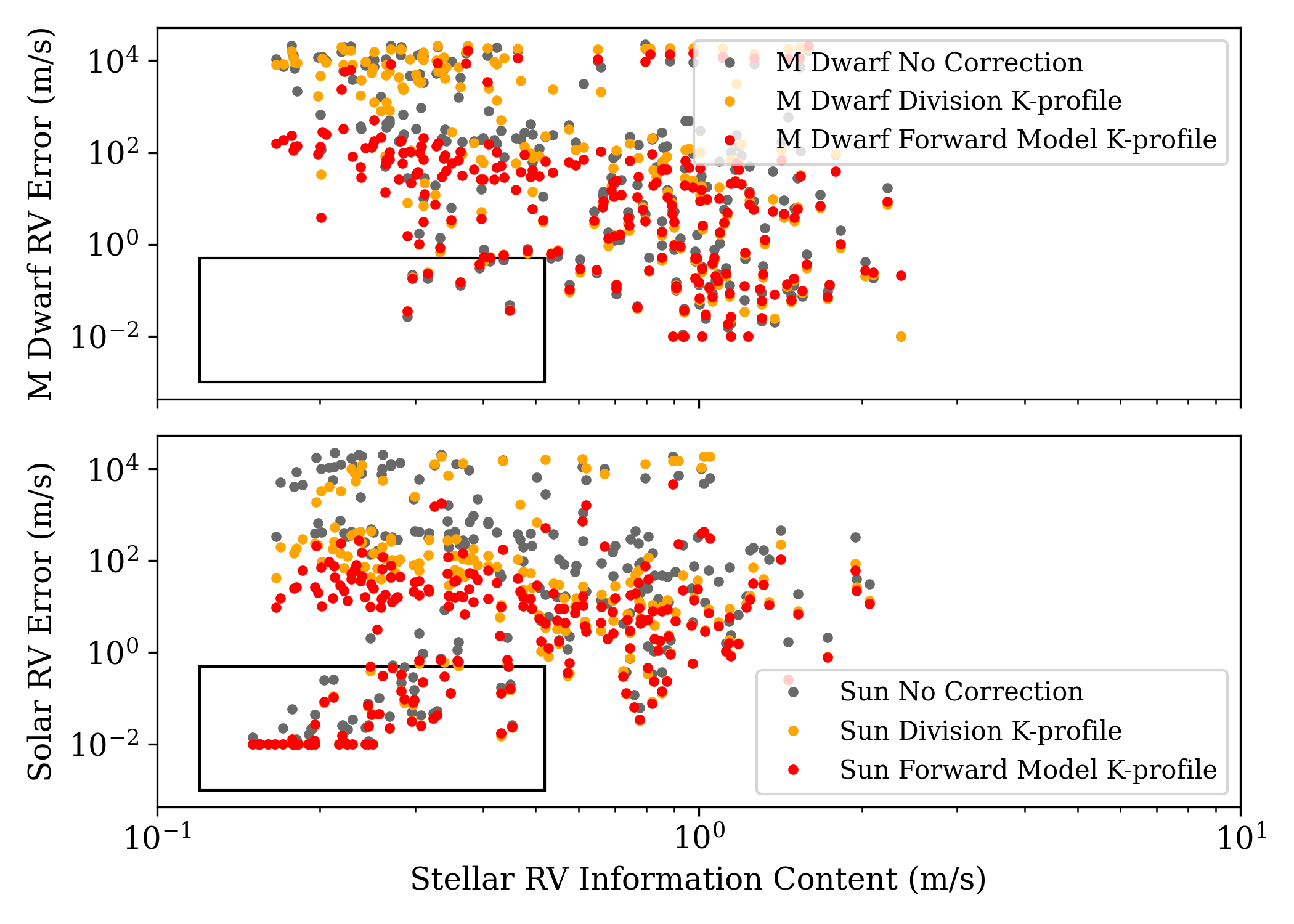}
\includegraphics[width=\linewidth]{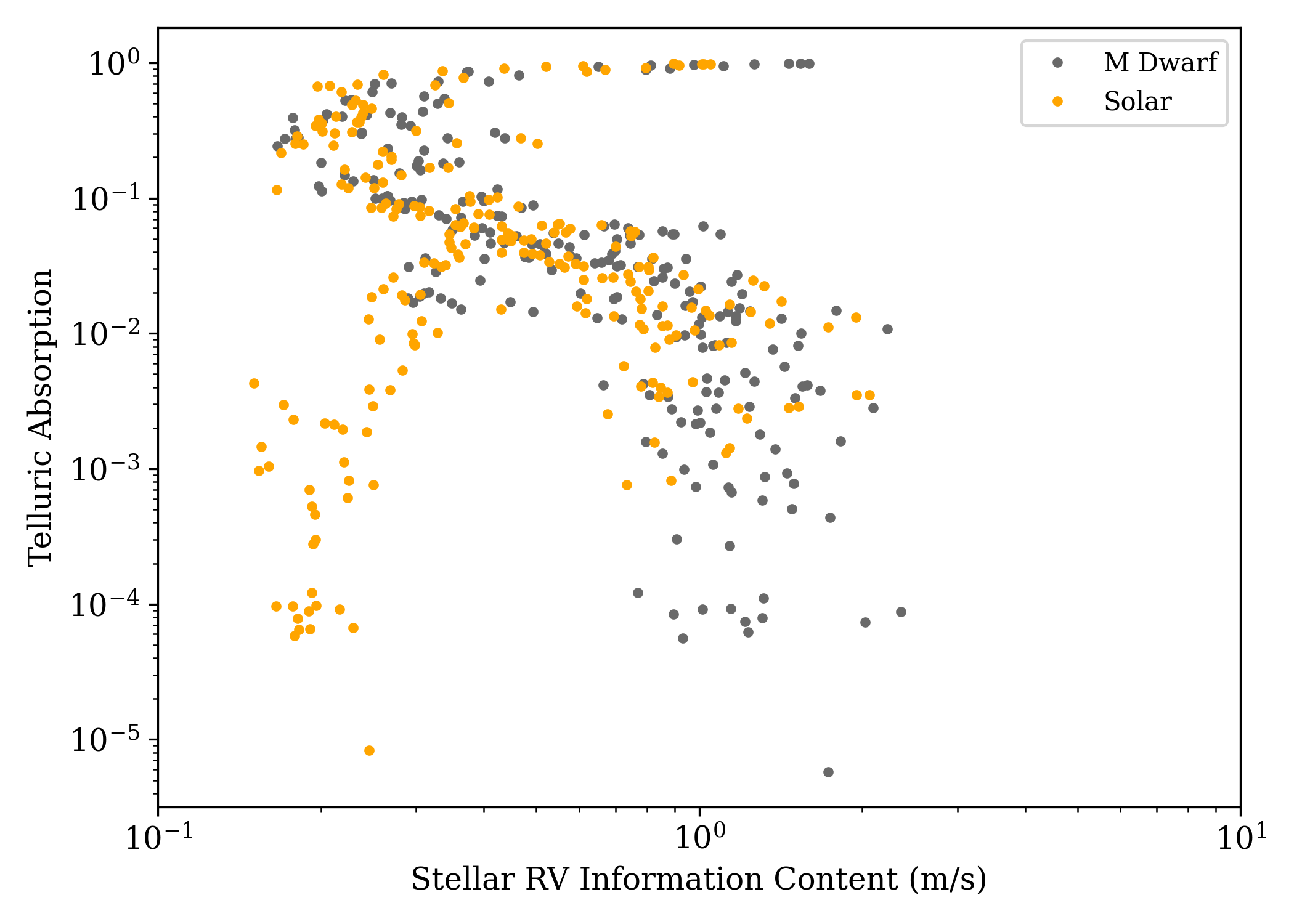}
\caption{Relationship between RV error and the stellar RV information content for both the M dwarf and Kurucz solar model. The top left plot shows the M dwarf RV error vs. the stellar RV information content and the bottom left plot shows the solar RV error vs. the stellar RV information content. In the left plot, the `no correction' results are shown in gray, the K-profile division results are shown in orange, and the K-profile Forward Model results are shown in blue. 
\label{fig:stelabspboth}}
\end{figure}

We study the comparison between an EarthFinder equivalent (EFE) instrument and other currently existing telescopes in Table~\ref{tab:rvinst}, broken down into visible and NIR arms as necessary \citep{plavchan2020earthfinder}. We can see that the EFE instrument has the widest wavelength range of any existing instrument, and the lowest photon-limited precision in the visible wavelength bands, and the EFE NIR arm outperforms or matches the currently available NIR instrument present in CARMENES, while going out further into the NIR than CARMENES. The first two rows presented in this table are reflected in Table~\ref{tab:resulttabboth} and the corresponding Figure~\ref{fig:RMSwave}. These results show the effectiveness of EFE in the NIR for M dwarf observations, which as discussed is also where the RVs are most affected by tellurics. We present the relationship between RV error and the stellar RV information content for both stellar models in Figure~\ref{fig:stelabspboth}. We also present the RV signal as a function of time for the M dwarf model only in Figure~\ref{fig:rvtime}. Further explanation for both can be found in Section~\ref{sec:comp}.

We also present further analysis of RV RMS as a function of photometric band in Table~\ref{tab:rvband}. As seen, the M dwarf RV RMS is significantly worse in the NIR than for the G star, specifically in the JHK bands. Similarly to Table~\ref{tab:resulttabboth} presented in this text, the results of Table~\ref{tab:rvband} are also presented in Figure~\ref{fig:banderror}. This figure contains all of the RV extraction methods discussed, and the horizontal lines mark the photon-noise weighted precision in each band, representing the last column in Table~\ref{tab:rvband}. Further explanation on Figure~\ref{fig:banderror} can be found in Section~\ref{sec:mitmethod}

\section{Discussion}
\label{sec:disc}
We present a discussion of comparison between M dwarf and G type, effectiveness to mitigation methods, caveats to our simulation, and iSHELL data comparisons.

\subsection{M \& G Dwarf Comparison}
\label{sec:comp}

In Table~\ref{tab:resulttabboth} we present a comparative summary of the average nightly RV RMS results for both the M dwarf and G type from W22 - with respect to wavelength range retrieved with each method described in Section~\ref{sec:method}: No correction, Division, Division with K-Profile, and Forward Modeling with the K-Profile. Between an M dwarf and a G dwarf, RV RMS differences are present in all wavelength bands present between 350 - 2500 nm. For an M dwarf (G star) in the visible band, a simulated EFE instrument as shown in Table~\ref{tab:resulttabboth} had an RV error RMS of 0.17 m/s (0.04 m/s) without correcting for tellurics, and 0.14 m/s (0.03 m/s) using Forward Modeling with K-profile for M dwarfs. In the NIR for the M dwarf (G star), there was an error of 2.2 m/s (2.4 m/s) without correcting for tellurics, and 1.6 m/s (1.2 m/s) with the K-profile Forward Modeling.  

We graphically represent the results from Table~\ref{tab:resulttabboth} in Figure~\ref{fig:RMSwave} for the M dwarf specifically; the full results are contained within the top two panels, while the bottom plot in Figure~\ref{fig:RMSwave} depicts a smoothed version of the results for clarity. As can be seen from this figure, the CCF division method with a perfectly known telluric line profile yields the best results, but is an unrealistic representation. However the forward modeling with K-profile, the most realistic method of simulation, yields the next best results and is the best realistic presented method between the other realistic scenarios of no correction or CCF division with K-profile. 

Based on the results of Table~\ref{tab:resulttabboth} and Figure~\ref{fig:RMSwave}, the M dwarf fares worse in the optical (380-900 nm) wavelength range than the G type star. This is expected, as the G type star has more stellar RV information content in the optical, where there are naturally fewer and shallower telluric lines (Figure~\ref{fig:spectel}). For the M dwarf, the peak in stellar RV information content appears in the red-optical, at approximately 700 nm, where the telluric lines are deeper and more frequent. This is explored in Figure~\ref{fig:stelabspboth}. In the NIR (900-2500 nm), both models return greater RV RMS as the telluric lines are deeply saturated in these wavelengths.

For M dwarfs, the tellurics more significantly impact PRV observations at NIR wavelengths. In the NIR, Forward Modeling with K-profile had slightly less RV RMS than in the visible with no correction. This is expected since M dwarfs are brightest in the NIR, with the highest RV information content in the red-optical \citep{reiners2020}. Thus for M dwarfs, as expected, the wavelengths with the most RV information are the most telluric-plagued. The stellar information content was calculated using the following equation from \citet{butler1996}, where $\sigma_{\bar{v}}$ is the error in mean velocity, dI/dV is the slope obtained by the spectrometer and $\epsilon_{I}$ is the fractional Poisson
error in the number of photons that arrive into a pixel $i$:

\begin{equation}
    \sigma_{\bar{v}} = \frac{1}{\sqrt{\Sigma(\frac{dI/dV}{\epsilon_{I}})^2}}
\end{equation}

For a larger amount of stellar RV information content (and thus lower in m/s), the RV precision (or the RV limit set by stellar information content) gets higher and vice versa, as illustrated in Figure \ref{fig:stelabspboth}. 
As can be seen from Figure~\ref{fig:stelabspboth}, the G star sun-like model has very low telluric impact at the wavelengths with the highest stellar information content. Conversely, the M dwarf has the lowest telluric impact where there is low stellar information content. 

\begin{deluxetable*}{cccccc}[t]
     \tablehead{
    \colhead{Stellar} & \colhead{Wavelength}& \colhead{No Correction} & \colhead{Division} & \colhead{Division} & \colhead{Modeling} \\
    \colhead{Type} & \colhead{Range (nm)} & \colhead{(m/s)} & \colhead{(m/s)} & \colhead{K-Profile (m/s)} & \colhead{K-Profile (m/s)}
    }
    \startdata
        M dwarf  & 380-900 & 0.167 & 0.125 & 0.120 & 0.140\\
         & 900-2500 & 2.158 & 0.636 & 1.622 & 1.636\\ 
        G dwarf & 380-900 & 0.035 &   0.021 &   0.021 &   0.032 \\
        & 900-2500 & 2.432 &   0.761 &   1.196 &   1.220 \\ 
    \enddata
    \caption{Summary of the scatter in the nightly RVs retrieved via each method for an M dwarf star and a G type star (calculated by W22). Column 1-2: Stellar Type, Wavelength Range (nm). Column 3-6: RV RMS for each method -- No Correction, Division, Division K-Profile, Modeling K-Profile, in m/s. These two wavelength ranges are for an EarthFinder Equivalent (EFE) instrument on the ground.}
    \label{tab:resulttabboth}
\end{deluxetable*}

\begin{figure}[h]
\centering
\includegraphics[width=\linewidth]{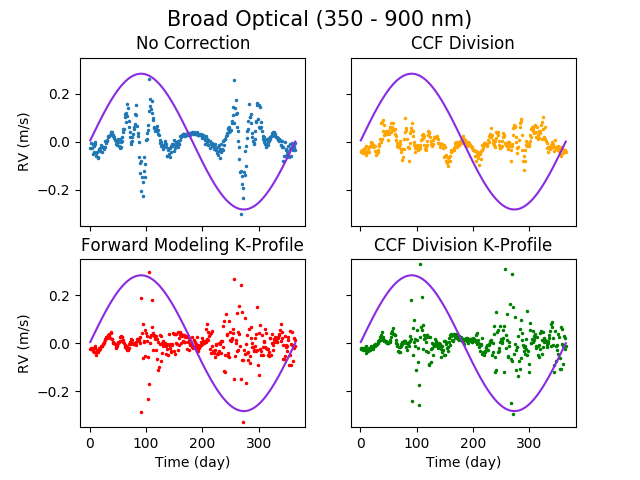}
\includegraphics[width=\linewidth]{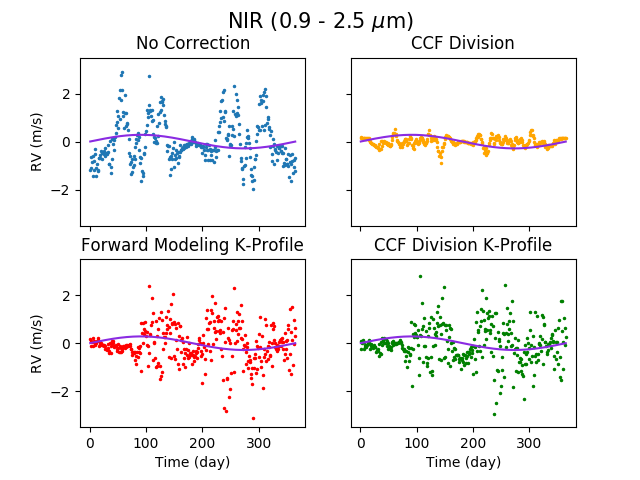}
\caption{RV signal as a function of time in units of days for the four RV extraction methods described in Section~\ref{sec:method} for the broad optical (left) and the NIR (right) for an EFE spectrograph. Shown in purple is an RV analog of Earth's signal (semi-amplitude of 25 cm/s). We assume SNR = 100 per pixel for R = 120,000. }
\label{fig:rvtime}
\end{figure}

In Figure~\ref{fig:stelabspboth}, the relationship between RV error induced by tellurics and stellar RV information content differs for the M dwarf and the G type star. Whereas for the solar model there is very low RV error in wavelength ranges with greater stellar RV information content, for the M dwarf the RV error is still high in areas with high stellar RV information content due to the tellurics present there as well. A larger stellar information content value means lower RV precision, and vice versa. In the right plot, this phenomena can also be seen in terms of telluric absorption line depth, with the M dwarf in blue and the solar model in yellow.
For the solar model, seen in the bottom panel within the black box, there are orders in the bottom left corner with little telluric absorption and high stellar RV information content. Notably, these orders are absent from the lower left corner box in the top plot with the M dwarf star. In the top plot, the lowest telluric RV error occurs where there are far less stellar lines, and consequently less precise RVs. This is also represented in terms of telluric absorption line depth in the right plot of Figure \ref{fig:stelabspboth}. As seen in this plot, where there is virtually no telluric absorption and a very high stellar RV information content and low stellar RV RMS, there are an abundance of solar model orders that are absent for an M dwarf. As expected, M dwarf RVs are more impacted by tellurics than G type stars, and we have quantified by how much in Table~\ref{tab:resulttabboth}.

We can also study the effectiveness of our extraction methods as a function of time. We show in Figure~\ref{fig:rvtime}, the RV signal as a function of time in units of days for all four RV extraction methods discussed are presented for the broad optical (350 - 900 nm) in the left panel, and the NIR (0.9 - 2.5 $\rm{\mu m}$) in the right panel for an EFE spectrograph. Also included in this figure is an RV analog of Earth's signal with a semi-amplitude of 25 cm/s for an M dwarf. This shows us precisely how the RVs in m/s vary as a function of time, wavelength range, and mitigation method. As can be seen, in the optical, the RV RMS is small compared to an Earth-like RV signal but when looking at the NIR, there is significant RV RMS when compared to the Earth-like RV signal. As expected and previously presented, this portrays the significant challenges posed by tellurics in the NIR region \citep{reiners2018}.

\begin{figure*}[t]
\centering
\includegraphics[scale=0.75]{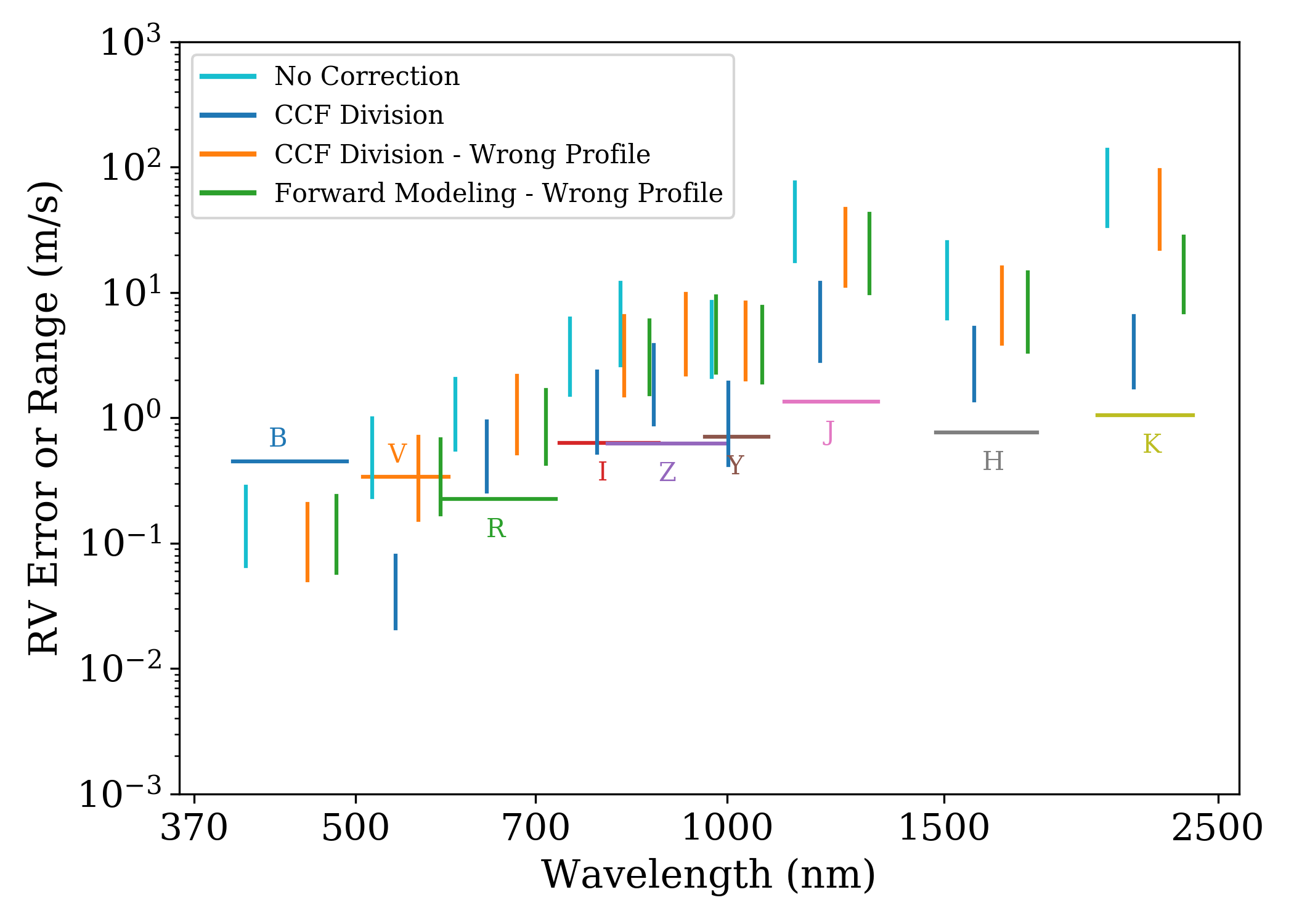}
\caption{Telluric induced RV errors for each photometric band. The horizontal lines mark the photon-limited precision for SNR=100 per pixel in each band. The lower ends of the vertical lines mark the RV RMS shown in Table~\ref{tab:rvband} and the upper ends of the lines mark the range of RV systematics (95\% percentile range) from the simulation methods listed in Table~\ref{tab:rvband}. 
\label{fig:banderror}}
\end{figure*}

\subsection{Effectiveness of Mitigation Methods}
\label{sec:mitmethod}

In Table~\ref{tab:rvinst}, we show the effectiveness of different mitigation methods for different ground-based instruments and their wavelengths. Looking at the last column, ``Photon-limited Precision (m/s)," in comparison to the first column, ``No Correction (m/s)" with instruments operating, we can see that tellurics did not considerably add to the error budget. Conversely, when we look at the instruments that operate in the NIR band, tellurics greatly contribute to the error budget in the optical band. Even division and Forward Modeling cannot reach close to the photon-limited precision, with a scatter of about 2 m/s. However, in general the three mitigation methods perform similarly. To explore this further with greater wavelength granularity, we peer into the different photometric bands to study the mitigation method effectiveness and photon-limited precision. In Table~\ref{tab:rvband}, we show the simulated RV RMS for multiple photometric bands. In the corresponding Figure~\ref{fig:banderror}, the horizontal lines show the photometric bands, the range of RV systematics (95$^{\rm th}$ percentile), and the RV standard deviation of any band are the two ends to every vertical line. This figure also illustrates that at SNR = 100 per pixel in the B and V bands, we are photon-noise limited; however tellurics generate larger RV RMS than the SNR = 100 photon noise would everywhere else. Conversely, this tells us the SNR at which telluric errors start to matter at each wavelength. Our analysis confirms a similar result for M dwarfs to that found in \citet{wang2019}, in that in the V band (roughly 500 - 600 nm), tellurics do not matter as much; they contribute errors on the order of $\sim$10 cm/s. Instead, the RV uncertainties for such an M dwarf in the V band is dominated by photon noise, as shown in Table~\ref{tab:rvband} and Table 1 in \citet{wang2019}.

\subsection{Additional Caveats to our Simulations}
\label{sec:caveat}

In our experimentation, we do not allow our code to fit for microtellurics during the first round of Forward Modeling when we fit for the optical depth for each species. Allowing fitting for microtellurics returns better results than only turning on species with deep enough telluric lines. During our runs in which we are fitting an incorrect line profile, fitting different orders independently gives our code more flexibility to compensate for the incorrect line profile for every molecular species. However, this approximation is unrealistic. We assumed a noise-free simulated spectrum, which allows for microtelluric line fitting down to incredibly small line depths, but in reality this is not possible. For instance, the water line depth cannot be recovered very well when they are obscured by photon noise, and the same logic applies with most microtellurics that are below typical noise levels. For this reason, we did not present the results of our trials with microtelluric fitting as our fiducial result.

\subsection{Empirically Fitting Tellurics with IRTF/iSHELL Data}
\label{sec:ishell}

Our K-profile approach is a good order-of-magnitude approximation to real observations where tellurics cannot be modeled perfectly, although it certainly does not represent the accurate and complete picture. In order to compare the idealistic Forward Modeling simulated scenario to real observations, we compare the simulated residuals in the Forward Modeling method with those obtained from iSHELL on the 3.0 m NASA Infrared Telescope Facility (IRTF) on Maunakea at K band wavelengths (2.2-2.5 $\mu$m). iSHELL is an R $\sim$ 80,000 spectrograph and has shown 3-5 m/s precision on year long timescales \citep{cale2019}. iSHELL also has a methane isotopologue gas cell (13CH$_4$) for wavelength solution and LSF characterization. As in \citet{cale2019}, we take high SNR ($>$ 100 per detector pixel near the continuum and blaze wavelength) spectra of Barnard’s Star and Forward Model them using an iterative approach to solve for the stellar template using the Python package {\tt pychell}. Relevant telluric species for the spectral range analyzed (2.28-2.41$\mu$m) are $\mathrm{H}_{2}\mathrm{O}$ and $\mathrm{CH}_4$, with water having the largest variability in opacity.

\begin{figure*}
    \centering
    \includegraphics[scale=0.5]{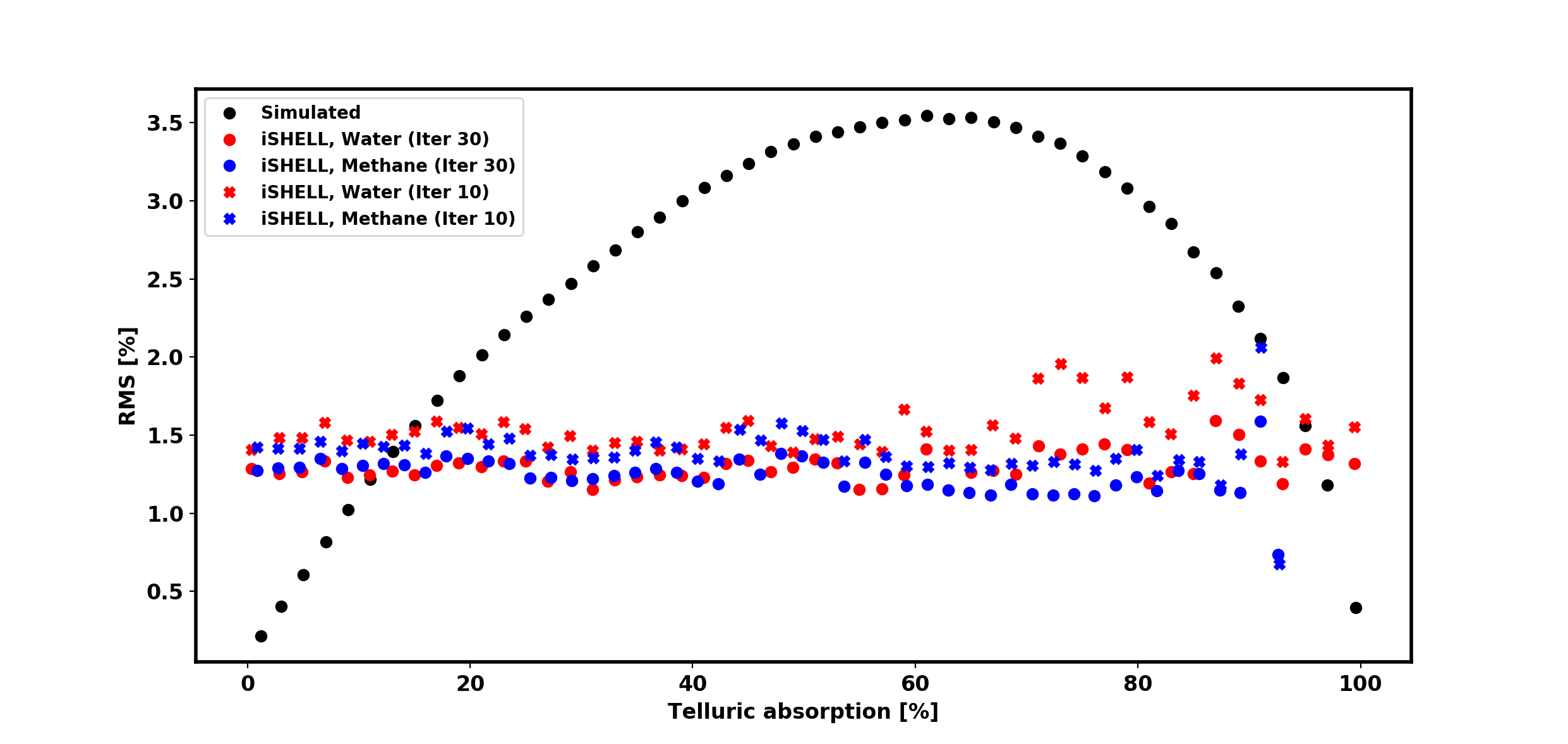}
    \caption{RMS of the flux residuals vs. telluric absorption depth from Forward Modeling iSHELL spectra at K-band (2.8-2.41 $\micron$) plotted with our simulations' chunks that match these wavelengths. Each point represents a residual flux RMS that corresponds to a small range of telluric absorption ($\approx \pm 1\%$).}
    \label{fig:rms_vs_tell_flux}
\end{figure*}

For a given echelle order, the two telluric input templates are first raised to their respective best fit optical depths from the Forward Model fit, and then convolved with a Gaussian LSF of width 0.1 $\angstrom$ to take into account the iSHELL line profile. Each modified telluric template is then linearly interpolated onto the unique wavelength grid for each observation. The modified water and methane templates are then separately binned using 50 equally spaced bins (in flux) with the minimum and maximum flux values as endpoints. We match the corresponding residuals to each bin for water and methane and compute an RMS. We directly sum residuals from (possibly) all echelle ($m=216-232$) orders and observations since each observation (and order) are at a similar S/N in the case of iSHELL, and are also identical at infinite photon SNR for the simulated observations. We perform similar steps on the simulated observations (across wavelength range 2284--2425 nm) using chunks 219--227 which approximately match the relevant iSHELL spectral range. We choose not to separate the methane and water analyses in this instance since the simulated observations treat the two species identically.

For the simulated residuals, we find the residuals are largest around 50\% telluric absorption suggesting the wings of the telluric lines suffer more from the ``mismatch'' of input line profiles, as shown in Figure~\ref{fig:rms_vs_tell_flux}. In the case of iSHELL, there is no clear correlation with telluric absorption depth for methane, but water possesses a higher RMS when absorption is $\gtrsim70$\%. Further, the mismatches are on average larger in the case of the simulated observations, indicating that our simulated induced line profile error is an overestimation of the real differences between the TAPAS templates and observations for Maunakea.

\begin{figure*}
    \centering
    \includegraphics[width = .8\linewidth]{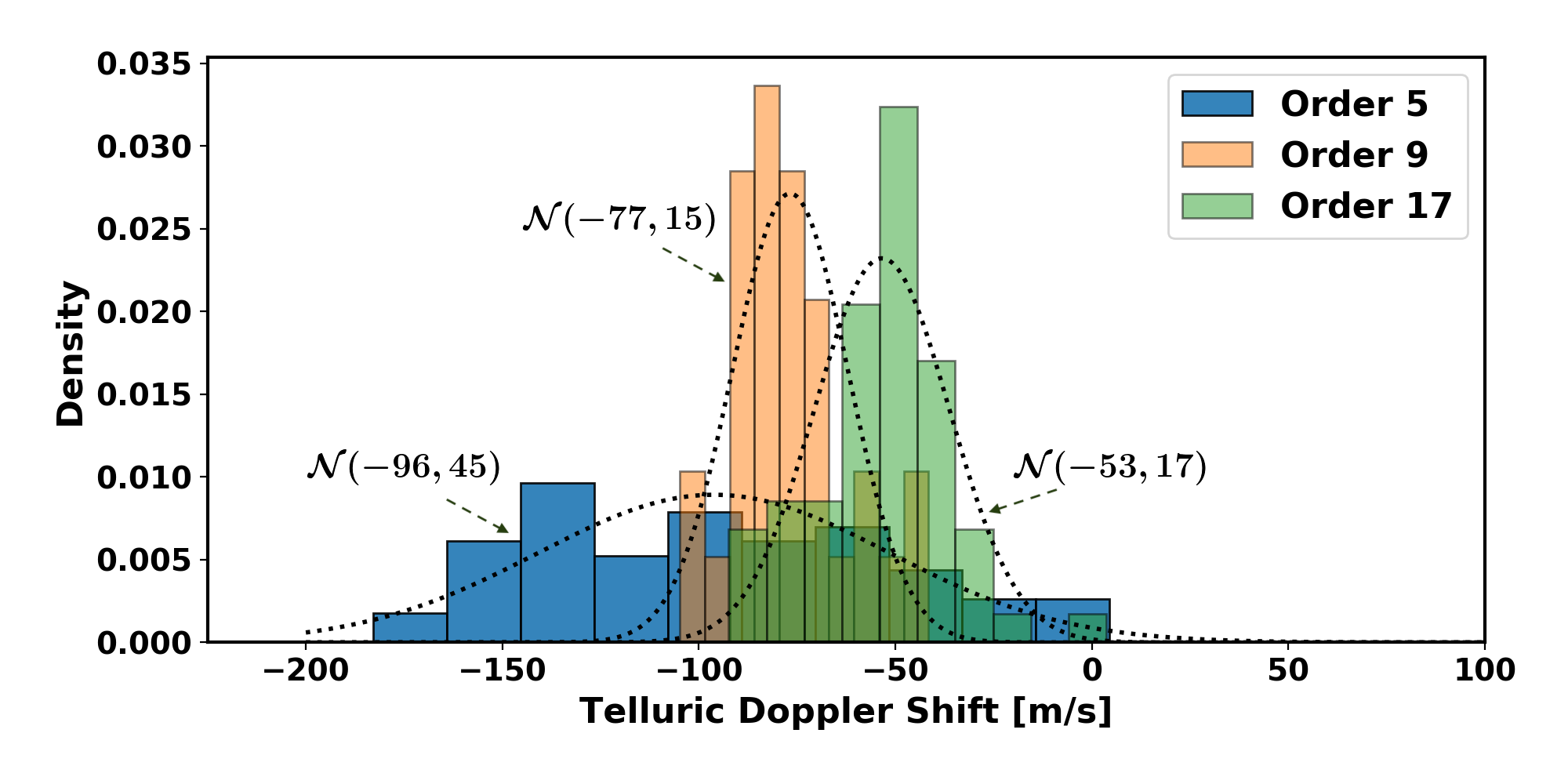}
    \caption{Histograms for the best fit telluric shift parameter for three iSHELL orders and corresponding best-fit normal distributions to the histograms (in units of \mos). Order 5 ($m=216$) has considerably more features from water than orders 9 ($m=220$) and 17 ($m=228$).}
    \label{fig:tell_shifts_ishell}
\end{figure*}

Obtaining a robust stellar template to compare iSHELL observations against \texttt{pychell} relies on isolating the intrinsic stellar spectrum present within the data by co-adding residuals in the barycenter rest frame where they are nearly coherent. This process is repeated up to 40 times for a given order before the RVs tend to stabilize \citep{cale2019}. Although other spectral features are forward modeled and are further decoherent in the barycenter frame, we do not always have the adequate barycenter sampling for a given target to fully remove any potentially remaining errors. Therefore, we can not assume that telluric features are fully excluded from the empirical stellar template.

In the case of iSHELL, we suspect the primary limiting factor in RV precision is the overall complexity of the Forward Model rather than tellurics. The solver iteratively optimizes 36 parameters, many of which are correlated in non-intuitive ways. The variable iSHELL blaze function requires local perturbations that we model with cubic splines, but these are commonly found to be correlated with the LSF Hermite coefficients and telluric specific parameters. This implies the blaze and LSF are working together to capture any mismatches in the telluric lines. The telluric Doppler shift is also commonly correlated with the stellar shift, although typically not until later iterations. See \citet{cale2019} for more details.

Other possibly limitations to this comparison between iSHELL and our simulations pertain to the telluric Doppler shift for iSHELL and the overall template ``zero-points''. The best fit telluric shift parameters for each iSHELL order follow a normal wavelength shift distribution; however the mean shifts across orders are not consistent, as shown in Figure~\ref{fig:tell_shifts_ishell}. Although we do not yet fully understand why this is the case, we suspect the higher density of highly variable water lines contribute to this effect. These offsets also hint that the telluric model may benefit from allowing for species-specific Doppler shifts.

\section{Conclusions \& Future Work}
\label{sec:conc&fw}

To summarize, tellurics add systematic error to our recovered RVs for every method of RV extraction simulated herein for an M dwarf star, with the exception of our control, perfect knowledge Forward Model. In the optical, tellurics add at least 0.1 m/s RMS error, and in the NIR tellurics can add in excess of 2 m/s, depending on the telluric mitigation method employed. By comparison, the expected RV semi-amplitude for an Earth-mass planet orbiting in the HZ of a 0.5 solar mass M dwarf star is $\sim$25 cm/s. This shows the detrimental effect of tellurics on M dwarf observations, especially since the most useful RV information content is found in red-optical and NIR wavelengths. This problem would be mitigated with a space-based instrument as shown in \citet{plavchan2020earthfinder}. In the absence of such an instrument, however, Forward Modeling partially mitigates the problem, bringing the telluric induced error down to $\sim$ 0.16 m/s. 

In future works, we will assess the dependence of our analysis on spectral resolution, the assumed telluric line profile error, and the impact of time-varying telluric absorption and airmass on exposures of finite duration. The latter will become increasingly important for $<$10 cm/s photon noise observations, as the exposure times will approach durations on the order of the atmospheric dynamic time-scales. Higher spectral resolution may give us the ability to mitigate telluric influence in a much better way by resolving the telluric line profile and providing crucial information about the CCF stellar line shape deformation from stellar activity. We will also extend the specific work of this project across the full M star sequence from M0 to M9.

\begin{acknowledgements}
We thank the referee for their thoughtful and constructive review of this manuscript. N. L. gratefully acknowledges support from an NSF
GRFP. N. L. is the lead author of the paper and performed the analyses and writing presented herein.
All iSHELL data were processed on ARGO, a research computing cluster provided by the Office of Research Computing, and the exo computer cluster, both at George Mason University, VA. 
PPP acknowledge support from NASA (EarthFinder Probe Mission Concept Study 16-APROBES16-0020,  Exoplanet Research Program Award \#80NSSC20K0251, JPL Exoplanet Exploration Program and Research and Technology Development) and the NSF (Astronomy and Astrophysics Grants \#1716202 and 2006517).

This work used the service provided by TAPAS, which is being supported by CNES (Centre National des Etudes Spatiales) and CNRS (Centre National de la Recherche Scientifique). TAPAS is a service maintained by ETHER data center. ETHER acknowledges for TAPAS the use of HITRAN data base and the LBLRTM radiative transfer code, the use of ECMWF data and the ETHER data center. We acknowledge useful discussions with Iouli Gordon and Larry Rothman.
\end{acknowledgements}
\bibliography{references}
\end{document}